%% file: main.tex
\documentclass[12pt]{scrartcl}
\addtokomafont{disposition}{\rmfamily}
\addtokomafont{subsection}{\large}
\addtokomafont{section}{\large}
\usepackage[utf8]{inputenc}
\usepackage[T1]{fontenc}
\usepackage[english]{babel}
\usepackage[top=25mm,bottom=25mm,left=25mm,right=25mm]{geometry}
\usepackage{lmodern}
\usepackage{microtype}
\usepackage{amsfonts}
\usepackage{amssymb}
\usepackage{amsmath}
\usepackage{amsthm}
\usepackage{color}
\usepackage{csquotes}
\usepackage{setspace}
\usepackage{booktabs}
\usepackage{multirow}
\usepackage[table,xcdraw]{xcolor}
\usepackage{paralist}

\newcommand\VRule[1][\arrayrulewidth]{\vrule width #1}
\usepackage{booktabs,siunitx}
\usepackage[flushleft]{threeparttable}
\usepackage[backend = biber, natbib=true,
  style=authoryear,ibidtracker=context,bibencoding=utf8,hyperref=true,doi=false,url=false,
  maxcitenames=2, maxbibnames=50, isbn = false, firstinits = true,
  uniquename=false, uniquelist=false,dashed=false,natbib=true,date=year]{biblatex}
\DeclareNameAlias{author}{last-first}
\DeclareNameAlias{editor}{last-first}
\DeclareNameAlias{translator}{last-first}

\renewbibmacro*{volume+number+eid}{%
  \printfield{volume}%
  \setunit*{\addnbspace}
  \printfield{number}%
  \setunit{\addcomma\space}%
  \printfield{eid}}
\DeclareFieldFormat[article]{number}{\mkbibparens{#1}}

\renewbibmacro{in:}{}
\AtEveryBibitem{%
  \ifentrytype{article}{
    \clearfield{note}}{}%
  \clearfield{eprint}%
}

\newcommand{\tab}[2]{%
  \begin{table}[htb]
    \centering
    \input{tables/#1}
    \caption[\protect\detokenize{#1}]{#2}
    \label{tab:#1}
  \end{table}

}

\usepackage{bbm}
\usepackage{bm}
\usepackage{hyperref}
\usepackage{listings}
\usepackage{xcolor}
\usepackage{textcomp}
\usepackage{graphicx}
\usepackage{subfigure}
\usepackage{longtable}
\usepackage{pdflscape}
\usepackage{tikz}
\usepackage{etoolbox}
\newcommand{\changefont}[3]{
	\fontfamily{#1} \fontseries{#2} \fontshape{#3} \selectfont}
\newcommand{\E}{{\changefont{cmss}{m}{n} \operatorname{\text{E}}}}

\definecolor{red3}{RGB}{205,0,0}
\definecolor{blue3}{RGB}{0,0,205}
\definecolor{green3}{RGB}{0,205,0}
\definecolor{hotpink4}{RGB}{139,58,98}
\definecolor{chartreuse4}{RGB}{69,139,0}
\definecolor{darkorchid4}{RGB}{104,34,139}
\definecolor{mediumorchid4}{RGB}{122,55,139}
\definecolor{darkgreen}{RGB}{0,100,0}
\definecolor{gold4}{RGB}{139,117,0}
\definecolor{darkred}{RGB}{139,0,0}
\definecolor{iseblue}{rgb}{0,0.35,0.62}
\definecolor{darkgoldenrod4}{RGB}{139,101,8}
\definecolor{brown4}{RGB}{139,35,35}
\definecolor{magenta}{rgb}{1, 0, 1}
\definecolor{darkgreen2}{rgb}{0, 0.5, 0}

\DeclareOldFontCommand{\bf}{\normalfont\bfseries}{\mathbf}
\DeclareOldFontCommand{\it}{\normalfont\itshape}{\mathit}

\newcommand{\refsec}[1]{Section~\ref{#1}}
\newcommand{\reftab}[1]{Table~\ref{tab:#1}}
\newcommand{\reftabs}[2]{Tables~\ref{tab:#1} and~\ref{tab:#2}}
\newcommand{\reffig}[1]{Figure~\ref{#1}}
\newcommand{\reffigs}[2]{Figures~\ref{#1} and~\ref{#2}}
\newcommand{\refequ}[1]{Equation~(\ref{#1})}
\newcommand{\refapp}[1]{Appendix~\ref{#1}}

\newcommand{\quantlet}[2]{\hfill
  \raisebox{-1pt}{\includegraphics[scale=0.05]{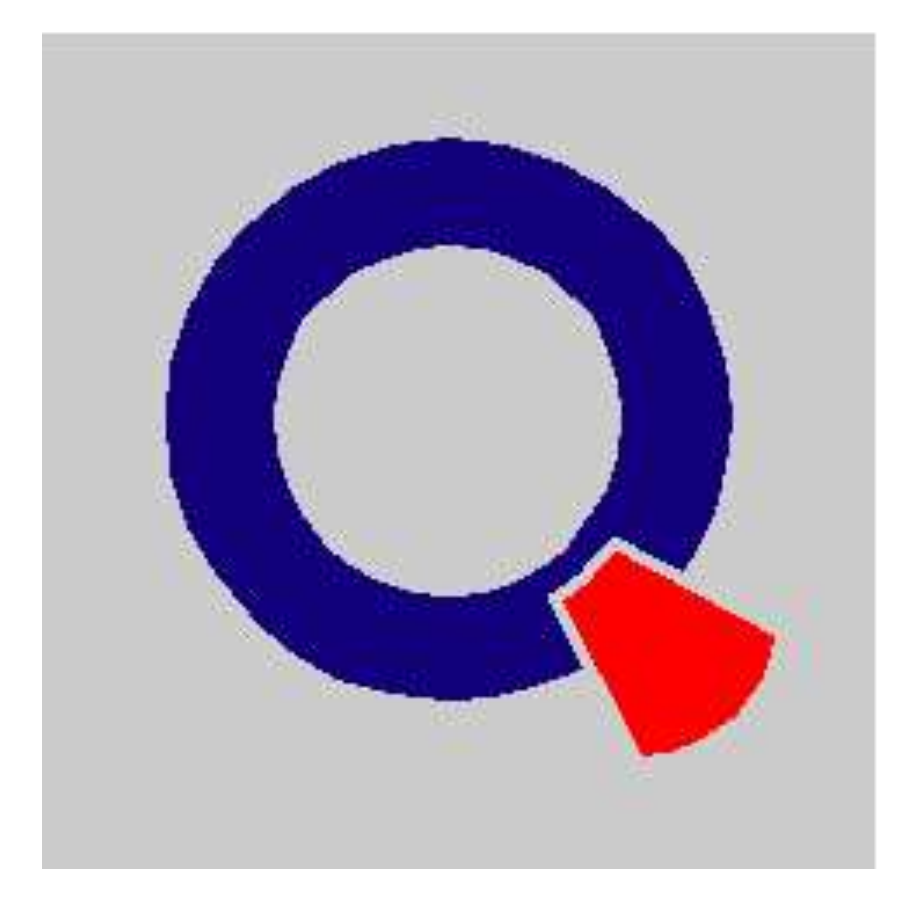}
    \href{https://github.com/QuantLet/CCP/tree/master/#1}{#2}}
}

\newcommand{\periodBeginEst}{2015-01-01}
\newcommand{\periodBegin}{2016-01-01}
\newcommand{\periodEnd}{2019-12-31} 
\newcommand{\periodRange}{from \periodBegin\ to \periodEnd}
\newcommand{\periodRangeAll}{\periodBeginEst\ to \periodEnd }
\newcommand{\XsecN}{52}

\newcommand{\trAmeans}{``TrA'' denotes only traditional, i.e., non-CC assets
  are included.}

\newcommand{\measuresAre}{The performance measures are final cumulative
  wealth (CW), the Sharpe ratio (SR), the adjusted Sharpe ratio (ASR), the
  certainty equivalent (CEQ), and turnover.}
\newcommand{\blacklineseparates}{the black line separates conventional assets
  (``TrA,'' upper yellow part of the spectrum) from cryptocurrencies (CCs,
  lower green-blue part of the spectrum)}
\newcommand{\withlibrolimits}{with a position limit of USD 10 mln (via the
  LIBRO approach)}
\newcommand{\descrstatvars}{$P_{10}$ and $P_{90}$ denote the first and ninth
  decile, respectively, ``Med'' the median, and ``SD'' standard deviation.}
\newcommand{\perfmeasures}{Performance measures for all investment strategies
  as well as benchmarks over the entire time period \periodRange}
\newcommand{\superiorperfred}{Highest results are highlighted in red.}
\newcommand{\diversifmeasures}{Measures of diversification for all investment
  strategies and a benchmark, over the entire time period \periodRange}
\newcommand{\diversifmeasuresexplained}{$DR^2$ denotes the squared
  diversification ratio, PDI the portfolio diversification index; all three
  measures are detailed in \refsec{divers_meas}.  \trAmeans\ Strategies are
  detailed in \reftab{list_strategies}.  Results without liqudity constraints
  (columns ``No const'') are contrasted with those when applying LIBRO with a
  threshold of USD 10 mln (column ``10 mln'').}
\newcommand{\perflegend}{with the following colour code:
  S\&P100, 
  \textcolor{magenta} {EW--TrA}, \textcolor{blue} {RR-MaxRet--TrA} and the
  corresponding \textcolor{red}{allocation strategy} from
  \reftab{list_strategies}}
\newcommand{\axesalignment}{Note that the date axes are aligned, but the
  wealth axes are not, due to large disperion in scales.}
\newcommand{\perfinCW}{\emph{Performance} in terms of cumulative wealth of
  portfolio strategies}

\providetoggle{long}

\title{\vspace*{-2ex} \LARGE Investing with Cryptocurrencies -- evaluating their potential for
  portfolio allocation strategies\footnote{Financial support from IRTG 1792
    ``High Dimensional Non Stationary Time Series,'' Humboldt-Universität zu
    Berlin, Czech Science Foundation under grant no.19/28231X and NUS FRC grant R-146-000-298-114 ``Augmented machine learning
    and network analysis with applications to cryptocurrencies and
    blockchains'', and the Yushan Scholar Program is gratefully acknowledged. The work of the authors is receiving support from the European Union’s Horizon 2020 training and innovation programme 
    ”FIN-TECH”, under the grant No. 825215 (Topic ICT-35-2018, Type of actions: CSA)}}

\author{Alla Petukhina \footnote{Humboldt-Universität zu Berlin, School of Business and Economics, Dorotheen Str.\ 1, 10117
    Berlin, Germany, tel: +49 (0)30 2093-99469, e-mail:
    \url{alla.petukhina@hu-berlin.de}} \and Simon Trimborn \footnote{Department of
    Mathematics, National University of Singapore, Risk Management Institute, 
    21 Heng Mui Keng Terrace, I3 Building 04-03, 119613 Singapore tel: +65
    6516-1245, e-mail: \url{simon.trimborn@nus.edu.sg}} \and Wolfgang Karl
  Härdle \footnote{Humboldt-Universität zu Berlin, IRTG 1792, Dorotheen Str.\ 1, 10117 Berlin,
    Germany; Wang Yanan Institute for Studies in Economics,  N114, Economics Building, Xiamen University
Xiamen, 361005 China; School of Business, Singapore Management University, 50 Stamford Road, Singapore 178899; 
    Faculty of Mathematics and Physics, Charles University, Ke Karlovu 3, 121 16 Prague, Czech Republic;  
    Department of Information Management and Finance, National Chiao Tung University, Taiwan, ROC, tel: +49 (0)30 2093-99592,
    e-mail: \url{haerdle@hu-berlin.de}} \and Hermann Elendner \footnote{Weizenbaum-Institut, Hardenbergstraße 32, 10623 Berlin, Germany, e-mail: \url{weizenbaum@elendner.net}} }

\begin{document}

\maketitle

\vspace*{-10pt}

\begin{abstract}
  \begin{center}
    {\bf Abstract}
  \end{center}

  Cryptocurrencies (CCs) have risen rapidly in market capitalization over the
  last years.  Despite striking price volatility, their high average returns
  have drawn attention to CCs as alternative investment assets for portfolio
  and risk management. We investigate the utility gains for different types of investors when they consider cryptocurrencies as an addition to their portfolio of traditional assets. We consider risk-averse, return-seeking as well as diversification-preferring investors who trade along different allocation frequencies, namely daily, weekly or monthly. Out-of-sample performance and diversification benefits are
  studied for the most popular portfolio-construction rules, including
  mean-variance optimization, risk-parity, and maximum-diversification
  strategies, as well as combined strategies.  To account for low liquidity
  in CC markets, we incorporate liquidity constraints via the LIBRO method.
  Our results show that CCs can improve the risk-return profile of
  portfolios.  In particular, a maximum-diversification strategy (maximizing the Portfolio
  Diversification Index, PDI) draws appreciably on CCs, and spanning tests
  clearly indicate that CC returns are non-redundant additions to the
  investment universe. Though our analysis also shows that illiquidity of CCs potentially reverses the results. 


\end{abstract}


\newpage


\doublespacing

\newpage
\section{Introduction}
\input{introduction}

\section{Literature review}
\label{lit_review}
\input{lit_review}

\section {Asset-allocation models}
\label{asset_models}
\input{asset_models.tex}

\section{Liquidity constraints with the LIBRO framework}
\label{liquidity_constr}
\input{liquidity_constr}

\section{Evaluating the performance of  portfolios}
\label{perform_meas}
\input{performance_measures}

\section{Data}
\label{data}
\input{data}

\section{Empirical results}
\label{emp_results}
\input{results}

\section{Conclusion}
\label{conclusion}
\input{conclusion}

\singlespacing

\newpage
\clearpage
\printbibliography

\newpage
\section{Appendix}
\label{appendix_math}
\input{appendix}

\end{document}

%% file: introduction.tex
Cryptocurrencies (CCs) have exhibited remarkable performance in the decade
since \textcite{Nakamoto_2008} invented the blockchain.  Accompanied by huge
inflows of capital into the market and strong swings in prices, CCs have
gained strongly in market value.  Accordingly, indices like CRIX
\citep[\url{thecrix.de}]{trimborn_crix_2016} were introduced to capture the market evolution
and provide a basis for ETFs.  Driven by these developments, cryptocurrency
markets became increasingly attractive to investors, who have started to
consider CCs as a novel class of alternative investments.  However, investors
differ with regard to their risk profiles, investment targets, individual
trading behaviors, and generally their diverse motives and preferences, and
thus the perspective to include CCs into financial portfolios raises a number
of questions:

\newcommand{\questionFour}{For whom is investing in the CC market valuable?
  Is the benefit derived from adding CCs to a portfolio dependent on the
  investor's objectives (e.g., return-oriented or diversification seeking)?}
\newcommand{\questionOne}{To which type of investor are CC investments most
  useful?  Only professional traders who rebalance their portfolio
  frequently, or also less actively trading retail investors?}
\newcommand{\questionTwo}{Should investors focus on one particular coin
  (e.g., Bitcoin), a selected few, or rather build a portfolio of a broad
  selection of CCs?}

\begin{enumerate}
\item \label{Question4}
  \questionFour
\item \label{Question1}
  \questionOne
\item \label{Question2}
  \questionTwo
\end{enumerate}

When an investor does decide to include CCs in the portfolio, further
questions arise about the choice of CCs for investment and their portfolio
weights:

\newcommand{\questionThree}{What exposure to each CC should be held in the
  portfolio?  How informative are past prices, how stable are positions when
  re-balancing the portfolio?  Do model-free strategies like equal-weighting
  provide reasonable results?}  
\newcommand{\questionFive}{Can these strategies be implemented in practice?
  In particular, are all CCs liquid enough for inclusion in an investment
  portfolio?  If not, how can investors still profit from promising CCs with
  little trading volume without exposing their portfolio too much to
  illiquidity?  Moreover, how is performance affected by honoring such
  portfolio restrictions?}  
\newcommand{\questionSix}{Overall, how do the properties of CC returns affect
  portfolios?  Is a certain type of portfolio-allocation method more suitable
  to manage and simultaneously exploit their properties?}

\begin{enumerate}
  \setcounter{enumi}{3}
\item \label{Question3}
  \questionThree
\item \label{Question5}
  \questionFive
\item \label{Question6}
  \questionSix
\end{enumerate}

While we review the literature extensively in \refsec{lit_review}, clearly
numerous studies have investigated the properties referred to in
Question~\ref{Question6}, and agree that CCs exhibit remarkably high average
realized returns by the standards of traditional financial assets\footnote{We
  do not compare CCs to derivatives, as they clearly constitute
  underlyings---in fact, a common complaint, albeit ignorant of their
  economic role, laments that CCs ``do not derive their value from any real
  asset.''  CC derivative markets still remain quite nascent.}---and
correspondingly high risk and uncertainty.  Not only is price volatility
high; also unfavorable properties obtrude, including frequent pricing bubbles
\citep{Fry_Cheah_2016, Hafner2018, Chen2019, Nunez_et_al_2019}, accumulation
of jumps \citep{Scaillet2018}, even evidence of price manipulation
\citep{Gandal_et_al_2018}.

At the same time, there is evidence of low correlations of CC returns with
those of traditional financial assets and other CCs.  Therefore, the high
risk of CC positions may be compensated by appropriate returns as well as
provide an opportunity to increase portfolio diversification.  Results to
that effect have been spearheaded by \textcite{briere_virtual_2015,
  eisl_caveat_2015}, the first to include Bitcoin (BTC) in a portfolio of
traditional assets, and subsequently bolstered by
\textcite{elendner_cross-section_2017}, who include a broad cross-section of
CCs, \textcite{chuen2017cryptocurrency}, who instead add CRIX, and lately
\textcite{Platanakis_Urquhart_2019, Akhtaruzzaman_Sensoy_Corbet_2019}, who
include Bitcoin in advanced portfolio optimization and find it enhances the
risk-return profile.

So evidence exists that CCs can be beneficial for investors
\citep{Pele_et_al_2020}.  However, taking the investor's perspective, we see
that while prior studies have covered crucially important aspects of
investing with CCs, the outlined questions \ref{Question4}--\ref{Question5}
remain fundamentally unanswered, for at least two reasons.  First, the key
result of a diversification benefit of CCs (or BTC) cannot be established
unless a broad set of non-CC alternative investments are included.  For
simplicity, in this paper we refer to all non-CC investments as ``traditional
assets,'' including alternative investments like gold or real estate, in
order to focus on the potential of adding CCs to well-diversified portfolios.
Second, it remains unclear which strategies can actually be implemented in
practice unless specific care is taken to address the frequently extremely
dry liquidity in CC markets.

The importance of addressing liquidity concerns is pinpointed by
\textcite{trimborn_investing_2017}, who introduce LIquidity Bounded
Risk-return Optimization (LIBRO) when considering a large sample of CCs added
to a portfolio consisting of the S\&P100, US bonds and commodities.  Given
the low liquidity of CCs as compared to traditional markets, LIBRO is
designed to protect investors from an inability to trade a CC in necessary
amounts due to low trading volume.

Against this background, we address the questions above by performing a
large-scale comparative investment-strategy study including \emph{both} a
broad range of traditional assets together with a broad cross-section of CCs.
Therein, we test the performance of an extensive set of common investment
strategies and thus consider different types of investors, while we employ
the LIBRO method to handle liquidity concerns.  We consider risk-oriented,
return-oriented, risk-return-oriented, and combined strategies; see
\reftab{list_strategies} for a full list of strategies under consideration.
We estimate extending-window and rolling-window approaches optimizations for
a sizable breadth of different common objective functions.  Finally, we
compare all strategies based on three different re-allocation frequencies,
namely daily, weekly and monthly, providing results for investors trading at
different frequencies.  To the best of our knowledge, we thus present the
broadest study on investing with CCs conducted so far.

Closest related to our paper are \citet{Akhtaruzzaman_Sensoy_Corbet_2019} and
\citet{Platanakis_Urquhart_2019}, both also studying the influence of CC
investment on optimal portfolio composition.  However, both include only
Bitcoin,\footnote{\citet{Platanakis_Urquhart_2019} do run a robustness test
  replacing Bitcoin with CRIX, acknowledging the importance of altcoins.
  Naturally, diversification \emph{across} CCs necessitates an optimization
  including their individual, distinct return series.} whereas we consider a
broad cross-section of \XsecN\ distinct CC price series.  Moreover, both
consider fewer traditional assets: industry portfolios (so equity only) in
the former paper, US equity and bond investments in the latter, plus
commodities in a robustness test.  In contrast, our set of traditional assets
is critically broader: first, as CCs trade globally, our international
approach includes equity returns for each of the 5 major economic areas
(Europe, USA, Japan, UK, China), as well as region-specific bond returns.
Second, we always include alternative investments, namely gold, real estate,
commodities, and the returns to FX trades between the five regions' fiat
currencies.  \reftab{list_assets} lists the traditional assets all our
portfolios include.  As we have pointed out, this emphatically goes beyond
quantitatively extending prior studies: unless \emph{both} a broad
cross-section of CCs \emph{and} of traditional assets are included, it
remains impossible to determine the magnitude of diversification benefits,
and more critically, also impossible to distinguish whether apparent benefits
of CCs are indeed present, or if CCs merely proxy for alternative assets.

Moreover, we cover a longer time horizon, and can thus include more than 2
years after peak CC prices; also, we consider more allocation strategies.
Most importantly, since we take the investor's perspective, we implement
LIBRO and contrast portfolios with weights that observe the liquidity
constraints with otherwise identical portfolios which do not: it turns out to
cricitally affect performance for several popular trading strategies.

Our study contributes to answering questions
\ref{Question4}--\ref{Question6}.  Spanning tests show that more than $50\%$
of the CCs considered can improve the efficient frontier of a portfolio
containing even our broad set of traditional assets.  We show that purely
risk-minimizing investors will optimally choose to mostly forego CC
investment; however, for investors with higher target returns their addition
seriously expands the efficient frontier.  Diversification-oriented investors
benefit most, even in terms of maximizing cumulative wealth.  We also
document that a lower rebalancing frequency (monthly) of the portfolios
generally enhances cumulated returns.  As mentioned, we confirm that several
CCs exhibit low liquidity, which can be tackled with the LIBRO approach.  Our
results highlight the severity of low-liquidity risk, and how analyses that
do not take this risk into account will compute investment returns that are
infeasible for any but the smallest personal portfolios.

The paper is organized as follows.  First, \refsec{lit_review} reviews the
related literature.  \refsec{asset_models} provides an overview of the
asset-allocation models under consideration, with a focus on connections
between them; therein \refsec{comb_models} explains the approach of model
averaging across investment strategies.  \refsec{liquidity_constr} reviews
the LIBRO method.  In \refsec{perform_meas} we explain the methodology for
comparing the performance of the models considered.  Our dataset of portfolio
components is described in \refsec{data}, and \refsec{emp_results} presents
the results of our analyses of out-of-sample performance of all portfolio
strategies with CCs and traditional assets.  We conclude in
\refsec{conclusion}.

Code to produce the results of this paper is available via
\url{www.quantlet.de}
\raisebox{-1pt}{\includegraphics[scale=0.05]{quantlet}\href{www.quantlet.de}}.

%% file: lit_review.tex
Modern portfolio theory builds on the CAPM \citep{markowitz_portfolio_1952,
Sharpe_1964, Lintner_1965}, both a theoretical equilibrium model and a
directly applicable statistical approach.  Yet, financial markets do not meet
its assumptions, so it lacks empirical accuracy.  Asset pricing and portfolio
optimization address this lack in one of two ways.

The first we call the \emph{financial-economics approach:} it follows
\citeauthor{Ross_1976}'s (\citeyear{Ross_1976}) arbitrage-pricing
theory\footnote{This approach puts the emphasis on the equilibrium model and
is thus often preferred by theorists.} which keeps the linear structure and
adds more factors to capture systematic patterns in returns.  Popularized by
\citeauthor{Fama_French_1992} (\citeyear{Fama_French_1992},
\citeyear{Fama_French_1993}), 
it was extended to factors for momentum \citep{Jegadeesh_Titman_1993,
Carhart_1997} or profitability and investment \citep{Fama_French_2015}.  In
principle, the approach renders portfolio optimization straightforward and
unidimensional: a portfolio is better, the higher its alpha (the intercept
after accounting for all factors' loadings).  In practice, the choice of
factors depends on the investment universe, and also for given asset classes
controversy remains about factors \citep[the ``zoo'' of][]{Cochrane_2011},
how to choose them \citep{Feng_Giglio_Xiu_2020}, even basic methodology
\citep{NovyMarx_2014}.

A strand of the literature on cryptocurrencies (CCs) is devoted to finding
and using factors in CC markets \citep[see][]{Liu_Tsyvinski_2019,
Elendner_2018, Hubrich_2017, Sovbetov_2018, Shen_Urquhart_Wang_2019};
however, in this paper we pursue the second approach.  

We term it the \emph{quantitative-finance approach,} due to its statistical
nature.  Its idea, in essence, says: if we can capture the (joint) return
distribution (and its dynamics) of all investable assets (and parameters
affecting them), then we can directly estimate portfolio weights to optimize
the desired performance metric.  Owing to the abundance of statistical
techniques for the variety of modelling choices and investment objectives,
this approach is most common in fund management.\footnote{An additional
benefit is how it links potential empirical shortcomings to insufficiently
captured statistical properties, offering remedy via more refined
methods. 
}  However, the easy customization has precluded a standard, unique approach.
A portfolio's optimal allocation thus depends crucially on three elements:
the investment universe, the investment strategy, and the investment
objective as defined by the metric of optimization.

Most fundamental is the determination of the \emph{investment universe.}  Our
paper focuses on its role by analyzing it for extensive sets of common
strategies and objective functions; concretely, on the potential of adding
CCs.  Historically, starting from
stocks\footnote{\citet{markowitz_portfolio_1952}.} and a risk-free interest
rate,\footnote{\citet{Sharpe_1964}.} the diversification benefits to adding
bonds \citep{Liu_2016}, foreign exchange
\citep{Kroencke_Schindler_Schrimpf_2013, Barroso_SantaClara_2015,
Ackermann_Pohl_Schmedders_2017}, real estate
\citep{Benjamin_Sirmans_Zietz_2001, AddaeDapaah_Loh_2005}, and commodities
\citep{Belousova_Dorfleitner_2012} including gold \citep{Hoang_et_al_2015}
have been established in the literature.\footnote{In fact, already
\citet{Roll_1977} had stressed the ``market portfolio'' ought to include
\emph{all} wealth.  Naturally, his critique has led to innumerous suggestions
for further asset classes that cannot all be part of our analysis, including
private equity \citep{Gompers_et_al_2010}, fine art
\citep{Mei_Moses_2002-AER, Campbell_2008}, or even fine wine
\citep{Fogarty_2010, Chu_2014}.}  We term \emph{all} these assets
``traditional investments'', and we include proxies for all of them in our
benchmark portfolio.  This breadth is key, as our goal is to investigate the
effect of including \emph{additionally} CCs.  Only the broad traditional
portfolio ensures we assess the diversification potential \emph{of CCs} as
investments: otherwise, CCs might merely substitute for other alternative
investments.

Considering CCs as investments contains subtle irony, as
\citet{Nakamoto_2008} pseudonymously introduced the blockchain as a
technology to serve as money,\footnote{More precisely, the intent was a
protocol with the emphasis on the tokens' role as medium of exchange, not as
stores of value.} not a profitable investment opportunity.  Thus, initially
doubt and debate shrouded the economic role\footnote{Some debate centered on
the question whether investments in CCs play an economic role similar to
gold: See \textcite{Dyhrberg_2016, Shahzad_et_al_2019} for affirmative views,
and \textcite{walther2018bitcoin} for a dissenting one.}  of CCs
\citep{Glaser_et_al_2014, Yermack_2015, Boehme_et_al_2015-JEP,
Baur_Hong_Lee_2017}.\footnote{Generally, mainstream economics has joined the
research effort on CCs deplorably late; it is now catching up, see for
instance \citet{Schilling_Uhlig_2019, Abadi_Brunnermeier_2019}.
Game-theoretic modelling has been more active, including \citet{Houy_2016,
Dimitri_2017, Caginalp_Caginalp_2019, Bolt_vanOordt_forthc}.}

However, after more than a decade of increasing demand, market
capitalizations, and trading volumes for a multiplying number of
CCs,\footnote{At the latest update of this writing, in 2020-Q2, the leading
dedicated information platform \url{coinmarketcap.com} records more than 5000
CCs traded at more than 21,000 markets, totalling a market capitalization
close to 250 billion USD (almost two thirds of which are due to Bitcoin),
with a 24-hour trading volume surpassing 150 billion USD.}  the recently
flourishing academic literature converges to the consensus that CCs
constitute investments, generally, and a distinct asset class in particular.
This literature can be categorized along two dimensions: first, which CC
investment is considered?  Only Bitcoin, also a fistful of other highly
visible CCs like Ethereum or Ripple, or a broad cross-section of tradable
CCs?\footnote{Note that the commonly reported thousands of CCs include mostly
such with extremely low liquidity: As of 2020-04-28, only 10 CCs exhibit
daily trading volumes exceeding 1 billion USD; volume below 100,000 USD
exists already among the top 200 CCs.}  Second, which portfolio allocations
are considered?  Only the CC(s), or also traditional markets?  If the latter,
only equity markets, or a broad range of traditional investments?

Regarding the first point, the literature started by investigating the
properties of the Bitcoin price process \citep{Kristoufek_2015,
Chu_Nadarajah_Chan_2015, Cheah_Fry_2015, Urquhart_2017, Blau_2018,
Bariviera_et_al_2017, Osterrieder_2017, Liu_Tsyvinski_2018}, establishing, in
essence, the presence of all critical properties of equity returns, yet often
up to an order of magnitude stronger: CCs exhibit exceptionally high mean
returns, and likewise volatility and drawdowns; returns also feature
extremely heavy tails and high heteroskedasticity.  Correlations with common
return series turn out extremely low to non-existent.

Given such low correlations, it is intuitive why including Bitcoin can
enhance a portfolio of traditional assets. 
Whereas the high riskiness leads to low portfolio weights for risk-averse
investors, an inclusion is beneficial as it improves diversification.

At the same time, exploding interest in blockchain led to an explosion in the
number of investable CCs.\footnote{Quick growth in the number of traded CCs
was mostly driven by the free-software nature of Bitcoin, allowing forks, and
to a lesser degree by development of new (sometimes blockchainless) CCs.}
This kickstarted investigations into the joint-return properties of a broad
cross-section of CCs \citep{elendner_cross-section_2017, Wang_Vergne_2017,
Brauneis_Mestel_2018, Zhang_Wang_Li_Shen_2018, Wei_2018}, which confirmed the
return characteristics of Bitcoin to be representative for the entire asset
class;\footnote{The reasons to consider CCs an asset class naturally go
beyond the similarity of their return processes; the major reason is that
their economic rationale differs decisively from all other asset classes, as
they constitute the only means to provide real resources to decentralised
apps.} yet generally so-called \emph{altcoins} exhibit still higher risk and
mean returns.  (Even more extreme were returns of Initial Coin Offerings,
ICOs, in particular during their peak in 2017---see, for instance,
\textcite{Adhami_Guidici_Martinazzi_2018, Momtaz2018, Momtaz2019,
Momtaz_2019}.  However, despite the important economic role of ICOs and STOs
(Security Token Offerings) as novel channels of venture-capital investment,
they are unsuitable for rules-based portfolio allocation, and hence fall
outside the scope of our paper.)

A key finding is that correlations are low even among CCs, as long as they
are no close substitutes or forks.  This implies a potential diversification
benefit from a broad basket of CCs \citep{chuen2017cryptocurrency}.
\textcite{Alessandretti2018}, optimizing CCs-only portfolios with LSTMs and
decision trees, also find enhanced return performance.

As one consequence, CC indices were developed: The CRIX
\citep{trimborn_crix_2016} captures the broad CC market movement with a
statistically optimized varying number of constituents; CCI30
\citep{rivin2018cci30} is a simple, close analogue to stock-market indices;
F5 \citep{Elendner_2018} is a momentum-factor-based,
transaction-cost-optimised basis for an exchange-traded portfolio; C20
\citep{crypto20} is an on-chain crypto-asset itself.  VCRIX \citep{Kim2019}
is a volatility index for option pricing.  A first paper on option pricing of
cryptos is \textcite{Hou_et_al_2020}.

The second key finding of cross-sectional analyses is that CCs beyond the
most prominent exhibit considerably low liquidity.  Portfolio calculations
ignoring liquidity might suggest trades which are impossible without extreme
price impact.  \citet{trimborn_investing_2017} introduce LIquidity Bounded
Risk-return Optimization (LIBRO) to account for illiquidity in CC portfolio
formation.  Since our focus is to evaluate the potential of adding CCs to
traditional portfolios, i.e., we take the investor's perspective, we provide
results both without and with the inclusion of LIBRO constraints.

In summary, the literature on CCs so far has solidly established potential
benefits of holding CCs in investment portfolios (foremost high returns and
low correlations), as well as certain difficulties (critically low
liquidity).  Yet open questions remain; prime among those whether CCs
``only'' proxy for alternative (non-CC) assets, or provide investment
opportunities that cannot be realized without CC positions.  We close this
gap by evaluating a wide range of common asset-allocation models with and
without CC positions.

%% file: asset_models.tex
Consider a matrix $X \in \mathbb{R}^{P \times N}$ of log returns of $N$
assets for $P$ days.  In our comparative analysis we rely on a moving-window
approach.  Specifically, we choose an estimation window of length $K = 252$
days (corresponding to the number of trading days in a calendar year).  We
investigate the performance of strategies for three rebalancing frequencies
$k$: monthly, with $k = 21$ days, weekly, with $k = 5$ days, and daily with
$k = 1$ day.\footnote{We also test strategies on extending windows as in
  \textcite{trimborn_investing_2017}; since the insights are similar, these
  results are not reported.}  For each rebalancing period $t$
($t = 1, \dots, T$, with $T$ the number of moving windows, defined as
$T = \frac{P-K}{k}$), starting on date $K + 1$, we use the data in the
previous $K$ days to estimate the parameters required to implement a
particular strategy.  These parameter estimates are then used to determine
the relative portfolio weights $w$ in the portfolio of risky assets.  Based
on these weights, we compute the strategy's return in rebalancing period
$t+ 1$.  This process is iterated by adding the $k$ daily returns for the
next period in the dataset and dropping the corresponding earliest returns,
until the end of the dataset is reached.  The outcome of this rolling-window
approach is a series of $P - K$ daily out-of-sample returns generated by each
of the portfolio strategies listed in \reftab{list_strategies}.  To simplify
notation, we omit the index $t$ for moving window or rebalancing period.

The traditional evaluation literature \citep[e.g.,][]{demiguel2009optimal,
  schanbacher2014combining} considers an investor whose preferences are
specified in terms of utility functions and fully described by the portfolio
mean $\mu_{P}$ and variance $\sigma_{P}$.
However, \textcite{merton1980estimating} showed that a very long time series
is required in order to receive accurate estimates of expected returns.  Due
to this high margin of error of expected-return estimates some authors,
including \textcite{haugen1991efficient}, \textcite{chopra1993effect} and
\textcite{chow2011survey}, suggest to rely only on estimates of the
covariance matrix as input of the optimization procedure.  Thus, investors
assume that all stocks have the same expected returns and under this strong
assumption the optimal portfolio is the global minimum-variance portfolio.
The minimum-variance portfolio strategy represents one of the so-called
risk-based portfolios, i.e., the only input used is the estimate of the
variance-covariance matrix.  In this paper we consider the most popular ones:
Maximum Diversification, Risk-Parity, Minimum Variance and Minimum CVaR
portfolio.  In \refsec{ind_models} we describe the individual strategies from
the portfolio-choice literature that we consider.  In addition totraditional
approaches, we consider a decision maker with risk preferences specified in
percentile terms, and portfolio construction based on higher moments of the
portfolio return-distribution, such as skewness and kurtosis.  Therefore, in
our comparative study we distinguish three groups of strategies:
return-oriented, risk-oriented (or risk-based, as in
\textcite{clarke2013risk}), as well as a tangency portfolio with Maximum
Sharpe Ratio (MV-S), which we categorize as a risk-return-oriented strategy.

Taking into account that the ranking of models changes over time, and
motivated by the fact that in many fields a combination of models performs
well \citep[see, e.g., ][]{clemen1989combining, avramov2002stock}, we also
extend our analysis to include the combination of portfolio models based on a
bootstrap approach inspired by \textcite{schanbacher2014combining,
  schanbacher2015averaging}.  The detailed methodology of combined portfolio
models is discussed in \refsec{comb_models}.
%

\subsection{Common asset-allocation models}
\label{ind_models}
In this section we review those models that we consider in the empirical
analysis.  We also discuss links between the strategies and give conditions
under which they are equivalent.  In general, when bringing the theoretical
models to the data, we employ in-sample moments of return distributions as
estimators of their theoretical counterparts; naturally, all evaluation then
concerns out-of-sample performance.  As subsequent prices provide new
information about assets' returns, all estimates are updated before any
rebalancing trades.

In all models we rule out short selling, a standard assumption in the CC
literature, given that---with the exception of bitcoin, for which futures are
traded since December 2017---taking short positions on CCs is at the very
least impractical, if not outright impossible.

\subsubsection{Equally-weighted portfolio}
The most na\"ive portfolio strategy sets equal weights (EW) for all
constituents: every asset gets a weight $w_i = 1/N$ for $i= 1, \dots, N$.  If
all constituents have the same expected returns and covariances, the EW
portfolio is mean-variance optimal.  However, there is no need for
assumptions or estimates regarding the distribution of the assets' returns to
implement EW.  Moreover, as \textcite{demiguel2009optimal} show, EW
allocations can actually perform well, in particular in settings of high
uncertainty, i.e., parameter instability---the model-free approach avoids
overfitting.  This is also the reason why the F5 crpto strategy builds on an
EW baseline benchmark.  

\subsubsection{Optimal mean-variance portfolio}
Many portfolio managers still rely on Markowitz' risk-return or mean-variance
(MV) rule, combining assets into an efficient portfolio offering a
risk-adjusted target return \citep{hardle_applications_2015}.  MV portfolios
are optimal if the financial returns follow a normal distribution (which,
generally, they do not), or if risk can be fully captured via volatility
(which, generally, it cannot).  Otherwise, MV serves as an approximation,
which in favor of tractability and convenience accepts the drawbacks widely
discussed in the literature: high portfolio concentration, i.e., high
portfolio weights for a limited subset of the investment universe, and high
sensitivity to small changes in parameter estimates of $\mu$ and $\sigma$,
see \textcite{jorion1985international}, \textcite{simaan1997estimation},
\textcite{kan2007optimal}.  In a Gaussian world, portfolio weights $w$ are
obtained by solving the following optimization problem:
 
\begin{equation}
  \begin{aligned}
    & \underset{w \in \mathbb{R}^p}{\text{min}}
    & & \sigma_{P}^2(w) \stackrel{\operatorname{def}}{=} w^{\top} \Sigma w\\
    & \text{\hspace{4mm}s.t.}
    & & \mu_{P}(w) = r_{T},\\
    & & & w^{\top}1_{N} = 1,\hspace{3mm} w_{i} \geq 0
  \end{aligned}
  \label{mv_opt}
\end{equation}
where
$\Sigma \stackrel{\operatorname{def}}{=} {\E}_{t-1}\{(X - \mu)(X -
\mu)^{\top}\}$ and $\mu \stackrel{\operatorname{def}}{=} {\E}_{t-1}(X)$ are
the sample covariance matrix and vector of mean returns respctively,
$\mu_{P}(w) \stackrel{\operatorname{def}}{=} w^{\top} \mu$, is the portfolio
mean and $r_T$ the \emph{target return,} ranging from minimum return to
maximum return to trace out an efficient frontier.  ${\E}_{t-1}$ is the
expectation operator conditional on the information set available at time
$t-1$.
         
We include three benchmark Mean-Variance portfolios in our analyses: first,
the global minimum variance portfolio (``MinVar'' in
\reftab{list_strategies}); second, the tangency portfolio (``MV-S''), and
third the portfolio with the highest in-sample return (``RR-MaxRet'').  In
our classification approach, MinVar is a risk-based decision rule, since it
is the most averse to risk and accepts the lowest target portfolio return.
At the opposite end of Markowitz' efficient frontier lies the
return-orientated RR-MaxRet portfolio, accepting any risk to choose the
(currently) highest possible reward.  In between these two endpoints, the
MV-S portfolio occupies middle-ground: it maximizes the Sharpe ratio
(\ref{SRatio}), in this way involving both risk and return estimation for the
portfolio construction.  We characterise MV-S as a risk-return-based
strategy.
    
\subsubsection{Optimal Conditional-Value-at-Risk portfolio}
A strong limitation of Markowitz-based portfolio strategies lies in the
assumption of Gaussian distributions of assets' log-returns.  Absent those,
for investors whose preferences are not fully described by a quadratic
utility funcion, variance or volatility is an insufficient risk measure,
leading the MV strategy to give a non-optimal portfolio composition.
Importantly, returns of CCs have even heavier tails as compared to those of
equities, as detailed in \textcite{chuen2017cryptocurrency} and
\textcite{elendner_cross-section_2017}.  The descriptive statistics of our
investment universe in \reffig{ReturnsDensity} and \reftab{DescrstatCC} in
\refapp{DescrStatApp} again provide strong evidence of this heavy-tailed
distributions for CCs.  Therefore, we include a strategy that accounts for
higher moments via Conditional Value at Risk (CVaR): we include a
Mean-CVaR-optimized portfolio as in \textcite{rockafellar2000optimization,
  krokhmal2002portfolio}.

For a given $\alpha < 0.05$ risk level, the CVaR-optimized portfolio weights
$w$ are derived as:
      
\begin{align}
  \underset{w \in \mathbb{R}^N}{\text{min}} \text{CVaR}_{\alpha}( w ),
  \hspace{2mm} \text{s.t. } \mu_{P}(w) = r_{T}, w^{\top}1_{p} = 1, w_{i} \geq 0,\\
  \label{cvar_opt}
  \text{CVaR}_{\alpha}(w) = -\frac{1}{1-\alpha}\int\displaylimits_{w^\top
  X\leq-\text{VaR}_{\alpha}(w)} w^\top X f(w^\top X| w)dw^\top X,
\end{align}
      
with $\frac{\partial}{\partial w^\top X} F(w^\top X|w) = f(w^\top X|w)$ the
probability density function of the portfolio returns with weights
$w$. $\text{VaR}_{\alpha}(w)$ is the corresponding $\alpha$-quantile of the
cumulative distribution function, defining the loss to be expected in
$(\alpha \cdot 100)\%$ of the times.
      
As for the MV portfolio, we construct the efficient frontier, from which to
derive the portfolios to add to our analyses.  As a risk-oriented strategy,
we add the MinCVaR strategy, minimizing the risk in terms of CVaR.  As far as
a return-oriented strategy is concerned, given our methodology, the maximal
expected return arises in the same way as in the maximum-return portfolio
(``RR-MaxRet'' in \reftab{list_strategies}), by investing in the riskiest
asset only.  Thus, we report this portfolio only as RR-MaxRet.

\subsubsection{Risk-parity portfolio (with equal risk contribution, ERC)}
One traditional risk-based portfolio strategy is based on the concept of risk
parity.  The underlying idea is to set weights such that each asset has the
same contribution to portfolio risk, see \textcite{qian2005financial}.
\textcite{maillard2010properties} derive properties of such portfolios and
rename them ``equal-risk-contribution'' (ERC) instruments.  The Euler
decomposition of the portfolio volatility
$\sigma_{P}(w)=\sqrt{w^{\top} \Sigma w}$ \citep{hardle_applications_2015}
allows to present it in the following form:
\begin{equation}
  \begin{aligned}
    \sigma_{P}(w)\stackrel{\operatorname{def}}{=}\sum_{i=1}^N \sigma_{i}(w) =
    \sum_{i=1}^N w_i\frac{\partial \sigma_{P}(w)}{\partial w_{i}},
  \end{aligned}
  \label{erc_opt}
\end{equation}
where $\frac{\partial \sigma_{P}(w)}{\partial w_i}$ is the marginal risk
contribution and
$\sigma_{i}(w) = w_i\frac{\partial\sigma_{P}(w)}{\partial w_{i}}$ is the risk
contribution of the $i$-th asset.  So, to construct the ERC portfolio, we
calibrate:
\begin{equation}
  \begin{aligned}
    \sigma_{i}(w)=\frac{1}{N} \quad \forall i 
  \end{aligned}
\end{equation}
The ERC portfolio can be compared to the EW portfolio: instead of allocating
capital equally across all assets, the ERC portfolio allocates the total risk
equally across all assets.  Consequently, if variances of log-returns were
all equal, the ERC portfolio would become identical to EW portfolio.  The ERC
portfolio is also comparable to the MinVar portfolio, which focuses on parity
of marginal contributions of all assets.

\subsubsection{Maximum-diversification portfolio (based on the Portfolio
  Diversification Index, PDI)}
Originally, the Maximum Diversification portfolio (MD) uses an objective
function introduced in \textcite{choueifaty2008toward} that maximizes the
ratio of weighted average asset volatilities to portfolio volatility or
diversification ratio as in \refequ{DRatio}.  In our study, instead of the
diversification ratio we maximize the Portfolio Diversification Index (PDI)
proposed by \textcite{rudin2006portfolio}.  It consists in assessing a
Principal Component Analysis (PCA) on the weighted asset returns' covariance
matrix, i.e., identifying orthogonal sources of variation.  In its original
form, PDI does not account for the actual portfolio weights, here we
incorporate weighted returns.  We optimize:
\begin{align}
  \underset{w \in \mathbb{R}^N}{\text{max}} \text{ PDI}_{P}(w), \quad
  \text{s.t. } w^{\top}1_{p} = 1, \quad w_{i} \geq 0 
  \label{MD_opt}
  \\
  PDI_{P}(w) {=} 2 \sum_{i=1}^N i W_{i} - 1,
  \label{PDI}
\end{align}
where $W_{i} = \frac{\lambda_i}{\sum_{i=1}^N\lambda_{i}}$ are the normalised
covariance eigenvalues $\lambda_{i}$ in decreasing order, i.e., the relative
strengths.  Thus, an ``ideally diversified'' portfolio, i.e., when all assets
are perfectly uncorrelated and $W_i = 1/N$ for all $i$, then $PDI ={N}$.  On
the contrary a $PDI \approx 1$ indicates diversification is effectively
impossible. Thus, in case of perfectly uncorrelated assets the MD portfolio
will be exactly the EW portfolio.  
The PDI summarises the diversification of a large number of assets with a
single statistic, and can compare the diversification across different
portfolios or time periods.  
 

%

\subsection{Averaging of portfolio models}
\label{comb_models}
Additional to individual allocation models, we also consider combinations of
models.  After all, every individual model is subject to estimation risk; the
idea of combining (or averaging) models in order to reduce such risk received
attention in various areas, and particularly in forecasting
\citep{avramov2002stock}.  Traditional model-averaging methods use
information criteria---like AIC or BIC---to identify relative shares of
models.  Across portfolio-allocation models the likelihood is unknown,
however, therefore we calculate model shares with the loss function $l$,
defined as 

\begin{equation}
  l(w) = w^{\top}\hat{\mu} - \frac{\gamma}{2}w^{\top}\hat{\Sigma}w.
  \label{ceq_loss}
\end{equation}

The parameter $\gamma$ reflects the investor's risk aversion, with $\gamma$
being large (small) for a risk-averse (risk-seeking) investor.  We use two
approaches to construct combined strategies: Na{\"\i}ve averaging of the
portfolio weights, as well as the combination method based on a bootstrap
procedure described in \textcite{schanbacher2014combining}.  However, in
order to account for possible time series dependencies at a daily frequency,
we apply the stationary bootstrap algorithm of
\textcite{politis1994stationary} with automatic block-length selection
proposed by
\textcite{politis2004automatic}. 

Consider a set of $m$ asset allocation models.  The corresponding portfolio
weights per model are given by $W = (w^1, \dots, w^m)$.  Relative shares of
(or beliefs in) individual models are $\pi = (\pi^1,\dots, \pi^m)$, such that
$\pi^{\top} 1_{m} = 1$. 
Then the asset weights for the combined portfolio are given by:

\begin{equation}
  w^{comb} =  \sum_{i=1}^m\pi^i w^i 
  \label{w_comb}
\end{equation}
 
The Na{\"\i}ve combination over all asset allocation models just assigns
equal shares, i.e., $\pi^{i}_{t} = \frac{1}{m}$ for all $i=1, \dots, m$.
 
The alternative, more sophisticated approach is to set the share
$\pi^{i}_{t}$ equal to the probability that model~$i$ outperforms all other
models.  We apply a bootstrap method to estimate these probabilities.  For
every period~$t$ we generate a random sample (with replacement) of $k$
returns using returns $X_{k(t-1)+1} \dots X_{k(t-1)+1+K}$, i.e., $K$-long
returns vectors of the $t-1$ rolling window.  We apply all $m$ asset
allocation models to these bootstrapped returns.  The procedure is repeated
$B$ times.  Let $s_{i,b} = 1$ if model $i$ outperforms in terms of the loss
function other models in the $b$-th bootstrapped sample, otherwise
$s_{i,b} = 0$.  The probability of model $i$ being best is then estimated as
\begin{equation}
  \hat{\pi}^{i}_{t} =  \frac{1}{B}\sum_{b=1}^{B}s_{i,b}
  \label{w_comb}
\end{equation}
where $B = 100$ is our number of independent bootstrap samples, and
$s_{i,b} = 1$ if model $i$ is the best model in the $b$-th sample.

\begin{table}
  \centering
  \scalebox{1}{
    \begin{tabular}{lll}
      \hline \hline
      {\bf Model} & {\bf Reference} & {\bf Abbreviation} \\
      \toprule
      \multicolumn{3}{l}{ \quad \it Model-free strategies} \\[1ex]
      Equally weighted	&	\textcite{demiguel2009optimal}	&	EW	\\
      \midrule
      \multicolumn{3}{l}{ \quad \it Risk-oriented strategies} \\[1ex]

      Mean-Variance -- min Var	&	\textcite{merton1980estimating}&	MinVar		\\
      Mean-CVaR -- min risk &	\textcite{rockafellar2000optimization}	&	MinCVaR	\\
      Equal Risk Contribution &	\textcite{maillard2010properties}&	ERC\\
      (Risk-parity)	&	&	\\

      Maximum Diversification& \textcite{rudin2006portfolio}&MD\\
      \midrule
      \multicolumn{3}{l}{ \quad \it Return-oriented strategies} \\[1ex]
      Risk-Return -- max return &	\textcite{markowitz_portfolio_1952} &	RR-MaxRet	\\
      \midrule
      \multicolumn{3}{l}{ \quad \it Risk-Return-oriented strategies} \\[1ex]
      Mean-Variance -- max Sharpe &	\textcite{jagannathan_risk_2003}	&	MV-S	\\

      \midrule
      \multicolumn{3}{l}{ \quad \it Combination models} \\[1ex]
      Na{\"\i}ve Combination      & 	\textcite{schanbacher2015averaging} & 	CombNa{\"\i}ve\\
      Weight Combination 	  & 	\textcite{schanbacher2014combining} & 	Comb\\

      \hline \hline
    \end{tabular}}
  \caption[List of strategies]{List and categorization of all asset
    allocation models we implement, including their abbreviations and
    references.}
  \label{tab:list_strategies}
\end{table}

%% file: liquidity_constr.tex
In this section, we review the LIBRO framework for portfolio formation, which
prevents too high portfolio weights for low-liquidity assets, by introducing
weight constraints in the portfolio optimization which depend on liquidity.

Liquidity, however, does not have a unique definition; different concepts and
measures abound.  \textcite{wyss_measuring_2004, Vayanos_et_al_2013} survey
the extensive literature on liquidity measures in equity markets; the
literature on CC liquidity is still scarce, with notable exceptions of
\textcite{Brauneis_et_al_2020, Scharnowski_2020}.  Due to the highly
fragmented market structure of CC exchanges (no dominant or central exchange
is trading all assets), we employ Trading Volume (TV) as our proxy for
liquidity.  TV is also the basis for the widely used \citetext{Amihud2002}
illiquidity measure, and proved suitable for the LIBRO methodology.  In
principle, alternative measures like the bid-ask spread would also be
applicable, as many exchanges report bid and ask prices; however, reliable
order-book data \emph{aggregated across} exchanges and for all CCs is
lacking.  TV, in contrast, is available for practically all CC markets, and
aggregated without problems.  For these reasons, we follow
\textcite{trimborn_investing_2017} and employ TV as our liquidity measure.
TV is defined as
\begin{equation}
  TV_{ij} = p_{ij} \cdot q_{ij},
\end{equation}
where $p_{ij}$ is the closing price\footnote{Technically, CC markets never
  close; the terminology ``closing price'' is still used in reference to the
  last price of a day, where days are customary defined on UTC time.} of
asset~$i$ at date~$j$, and $q_{ij}$ is the volume traded at date~$j$ of
asset~$i$.  The liquidity of asset~$i$ in period~$t$ can then be measured
with the sample median of trading volume,
\begin{equation}
  TV_i = \frac{1}{2} (TV_{i,up} + TV_{i,lo}), 
\end{equation}
where $TV_{i,up} = TV_{i,\lceil \frac{l + 1}{2}\rceil}$ and
$TV_{i,lo} = TV_{i,\lfloor \frac{l + 1}{2}\rfloor}$.

Define $M$ as the total amount invested in all $N$~assets, so that $M w_i$
denotes the market value held in asset~$i$.
\textcite{trimborn_investing_2017} formulate the constraint on the weight of
asset $i$ as
\begin{align}
  M w_i \leq TV_i \cdot f_i,
  \label{equ_w_bound}
\end{align} 
where $f_i$ captures the speed with which an investor intends to be able to
clear the current position in asset~$i$ via multiples of TV.  For example,
$f_i = 0.5$ implies the position in asset $i$ must not exceed 50\% of median
trading volume.  It results in a boundary for the weight on asset~$i$ as
\begin{align}
  w_i \leq \frac{TV_i \cdot f_i}{M} = \widehat{a}_i.
\end{align}

The beauty of this approach lies in its ease to include it into any portfolio
optimization.

%% file: performance_measures.tex
While \refsec{asset_models} presents the set of common asset-allocation
models we implement, no unique metric exists to evaluate and compare them.
In order to draw conclusions about the effect of adding CCs to broadly
diversified portfolios, we pursue three dimensions: First, we calculate a
range of widely used performance measures in \refsec{perf_meas}.  Second, in
\refsec{significance_test} we run direct tests for differences between
strategies on a pair-wise basis.  Third and finally, in \refsec{divers_meas}
we address the diversification effect of CCs directly by calculating three
well-known measures of portfolio concentration.

\subsection{Performance measures}
\label{perf_meas}
To assess the performance of the investment strategies we consider as it
develops over time, we employ the following five common performance criteria
widely used in literature, as well as by practitioners.  Performance measures
are computed based on the time series of daily out-of-sample returns
generated by each strategy.

First, we measure the \emph{cumulative wealth (CW)} generated by each strategy~$i$
\begin{equation}
  W_{i,t+1} = W_{i,t} + {\hat{w}_{i,t}^{\top}X_{t+1}}, 
  \label{cumret}
\end{equation}
starting with an initial portfolio wealth of $W_0 = \$1$. Cumulative wealth,
while naturally of high interest as a measure of performance achieved over
the period considered, is not sufficient to rank our allocation approaches.
Therefore why we also compute two traditional measures of risk-adjusted
returns: the Sharpe ratio, and the certainty equivalent.  Moreover, we
provide the Adjusted Sharpe Ratio (ASR) in order to address the MinCVaR
strategy and the non-Gaussian nature of the return distributions.

The \emph{Sharpe Ratio} (SR) of strategy~$i$ is defined as the sample mean of
out-of-sample excess returns (over the risk-free rate), scaled by their
respective standard deviation.  This definition presumes an unambiguous
risk-free rate, inexistent in the global context of CCs.  Fortunately, our
sample period is characterized by most of the global economy at or very close
to the zero lower bound on interest rates; so we can sidestep the question by
implicitly setting the riskless rate to 0 and defining
\begin{equation}
  \widehat{SR_i} = \frac{\hat{\mu_i}}{\hat{\sigma_i}^2}.
  \label{SRatio}
\end{equation}

The \emph{Certainty Equivalent} (CEQ) captures, for an investor with a given
risk aversion $\gamma$, the riskless return that said investor would consider
of equal utility as the risky return under evaluation.  For the case
$\gamma = 1$, it is equivalent to the close-form solution of
\textcite{markowitz_portfolio_1952} portfolio optimization problem in
\refequ{mv_opt}.
\begin{equation}
  \widehat{CEQ_{i,\gamma}} = \hat{\mu_i}-\frac{\gamma}{2}\hat{\sigma_i}^2 
  \label{ceq}
\end{equation}
While there is debate about the risk-averion coefficient best describing
investors going back to \citet{Mehra_Prescott_1985}, we argue that current CC
investors are unlikely to be characterized by extremely high risk aversion,
and calculate the CEQ in the empirical part of our paper with a $\gamma$ of
$1$.  As can be noted, the CEQ corresponds to the loss function $l$ defined
in \refequ{ceq_loss}.

The CEQ and in particular the SR are more suitable to assess of strategies
when assets exhibit normally distributed returns.  To address this drawback,
\textcite{pezier2006relative} propose the \emph{Adjusted Sharpe Ratio} (ASR).
ASR explicitly incorporates skewness and kurtosis:
\begin{equation}
  \widehat{ASR_i} = \widehat{SR_i} \left[1 +
    \left(\frac{S_i}{6}\right)\widehat{SR_i} -
    \left(\frac{K_i}{24}\right) \widehat{SR_i}^2 \right]
  \label{SRatio}
\end{equation}
where $SR_i$ denotes the Sharpe Ratio, $S_i$ the skewness, and $K_i$ the
excess kurtosis of asset~$i$.  Thus, the ASR accounts for the fact that
investors generally prefer positive skewness and negative excess kurtosis, as
it contains a penalty factor for negative skewness and positive excess
kurtosis.

To assess the impact of potential transaction costs associated with asset
rebalancing, we also calculate two measures for turnover.  \emph{Portfolio
  turnover} is computed to capture the amount of trade necessary on
rebalancing dates as
\begin{equation}
  {TO_i} = \frac{1}{T-K}{\sum_{t=1}^{T-K}}{\sum_{j=1}^N}\vert
  \hat{w}_{i,j,t+1}-\hat{w}_{i,j,t+}\vert
  \label{turnover}
\end{equation}
where $w_{i,j,t}$ and $w_{i,j,t+1}$ are the weights assigned to asset~$j$ for
periods~$t$ and $t+1$ and $w_{i,j,t+}$ denotes its weight just before
rebalancing at $t+1$.  Thus, we account for the price change over the period,
as one needs to execute trades in order to rebalance the portfolio towards
the $w_t$ target.  High turnover will imply significant transaction costs;
consequently, the lower Turnover of a strategy, the better it performs.

\emph{Target turnover,} the second turnover-related measure,
  captures the amount of change in target weights between two consecutive
  rebalancing dates as
\begin{equation}
  {TTO_i} = \frac{1}{T-K}{\sum_{t=1}^{T-K}}{\sum_{j=1}^N}\vert
  \hat{w}_{i,j,t+1}-\hat{w}_{i,j,t}\vert
  \label{target-turnover}
\end{equation}
In contrast to \refequ{turnover}, here the difference between weights spans
the time interval of one rebalancing period, instead of the (conceptually
infinitesimal) duration of rebalancing trades.  Therefore, the realized price
paths of the assets affect the measure only insofar as they lead to different
parameter estimates and thus a revision in target weights.  The difference
between the two turnover measures is best illustrated by considering the EW
strategy: it may require high turnover to return to exactly equal weights per
asset every rebalancing date; yet by definition it will never exhibit and
target turnover.

\subsection{Testing for performance differences between strategies}
\label{significance_test}

To test if strategies are significantly different from each other, we provide
the $p$-values of pairwise tests.  The common approach by
\textcite{jobson1981performance} 
is widely used in the performance evaluation literature \citep[e.g., also
in][]{demiguel2009optimal}.  However, this test is not appropriate when
returns have tails heavier than the normal
distribution. 
Therefore, as a testing procedure we rely on the \textcite{ledoit2008robust}
test with the use of robust inference methods.  We test for difference of
both CEQ and SR, and report results for the HAC (heteroskedasticity and
autocorrelation) inference version.  The procedure is described in
\refapp{SRdiff}.


\subsection{Measuring diversification effects}
\label{divers_meas}

To evaluate portfolio concentration and portfolio diversification effects, we
calculate three measures:
\begin{inparaenum}[\it a)]
\item the \emph{Portfolio Diversification Index} (PDI) as in introduced in
  \refequ{PDI},
\item Effective N as introduced by \textcite{strongin2000beating}, and
\item the Diversification Ratio.
\end{inparaenum}

\emph{Effective N} is defined as
\begin{equation}
  N_{\textit{eff}}(w_t)=\frac{1}{\sum_{j=1}^N{w^2_{j,t}}}
  \label{Neff}
\end{equation}
with $j = 1, \dots, N$ indexing assets.  Effective N varies from 1 in the
case of maximal concentration, i.e., the portfolio entirely invested in a
single asset, to $N$---its maximum achieved by an equally-weighted portfolio.
The design of Effective N is related to other traditional concentration
measures, e.g., the Herfindahl Index, the sum of squared market shares to
measure the amount of competition.  Effective N can be interpreted as the
number of equally-weighted assets that would provide the same diversification
benefits as the portfolio under consideration.

The \emph{diversification ratio,} suggested by
\textcite{choueifaty2011properties}, measures the proportion of a portfolio's
weighted average volatility to its overall volatility:
\begin{equation}
  DR(w_t) = \frac{w_t^{\top}\sigma_t}{\sqrt{w_t^{\top} \Sigma_t
      w_t}}=\frac{w_t^{\top}\sigma_t}{\sigma_{P,t}(w_t)}
  \label{DRatio}
\end{equation}
Thus, the diversification ratio has the form of the Sharpe Ratio in
\refequ{SRatio}, with the sum of weighted asset volatilities replacing the
expected excess return.  In case of perfectly correlated assets, the DR
equals 1; in contrast, in a situation of ``ideal diversification,'' i.e.,
perfectly uncorrelated assets, $DR = \sqrt{N}$.  Hence, in our empirical
study we report the results for $DR^2$, for two reasons: First, to make it
comparable to the other two used metrics, and second, because
\textcite{choueifaty2008toward} demonstrate that for a universe of $N$
independent risk factors, the portfolio that weighted each factor by its
inverse volatility would have a $DR^2$ equal to $N$.  Hence $DR^2$ can be
viewed as a measure of the effective degrees of freedom within a given
investment universe.

%% file: data.tex
For the empirical analysis, we collect daily price data on a sample of CCs
and traditional financial assets (including alternative investments) over the
period \periodRangeAll\ (1304 daily log-returns).  CC prices are provided by
CoinGecko, data for traditional assets is acquired from Bloomberg.  Many CCs
were established only after January 2015, or ceased to trade prior to the
period we study.  Since investors who apply rules-based optimization
techniques usually only consider assets with sufficient price histories, we
require CCs to have a continuous return time-series over the period of our
study in order to be included.  By excluding coins that did not already
circulate in January 2015, went extinct before December 2019, or have only
patchy price series, we effectively focus on solid CCs, of interest to
investors considering positions in this novel asset class.\footnote{We also
  run our entire analysis for a sample period extending until end of
  December, 2017.  For this shorter period, 55 CCs fulfil our criteria, and
  with minor exceptions only for combined strategies, all our results remain
  qualitatively unchanged.}  We also sidestep ICOs.  Hence, our final data
sample for portfolio construction includes \XsecN\ CCs next to 16 traditional
assets.  In order to cover 3 different reallocation frequencies (daily,
weekly, monthly), we calculate with daily, weekly and monthly return series
for all assets treated equally.

We employ a rolling-window approach for the portfolio construction.  The
initial portfolio weights are determined from estimations based on the first
year (2015), after which we `roll' through the dataset by estimating new
portfolio weights at the reallocation frequency.  Depending on the employed
frequency approach, this adds one day, week, or month of data to the
estimation set and leaves out the oldest day, week or month of data, in order
to capture potentially time-varying parameters.\footnote{As a robustness
  test, we also calculate with extending windows, where no historical data is
  dropped and only new observations added as they become observable.  The
  results are qualitatively the same.}

To evaluate the performance of each of the strategies we consider, our
research question studies the effects of including CCs \emph{as an addition
  to} classical, well-diversified portfolios.  Therefore, our investment
universe always includes 16 traditional assets from 5 asset classes: equity,
fixed-income, fiat currencies, commodities, and real estate.  Since CCs are
global in nature, our traditional assets cover the 5 main economic areas
around the globe (Europe, USA, UK, Japan, China).  In this way, the asset
space is sufficiently broad to allow diversification without CCs, ensuring
any relevance of CCs we find is genuine, and at the same time is still narrow
enough to allow us to add each CC individually as an asset without leading to
high-dimensionality issues in covariance estimation.  The full list of
traditional constituents of the investment universe is provided in
\reftab{list_assets}.  \reftabs{DescrstatCC}{DescrstatIND} in
Appendix~\ref{DescrStatApp} report summary statistics of all constituents
considered in our empirical study.

The main properties of our data correspond to the findings of the prior
literature, e.g., \textcite{chuen2017cryptocurrency}: CCs outperform
traditional asset classes in terms of average daily realised returns, their
returns exhibit higher volatility, with means mostly positive while the
medians are mostly negative, positive movements occur less frequently than
negative ones, but with higher magnitudes (absolute values of minima and
lower deciles are less than of maxima and higher deciles for the majority of
CCs).  Correlation analysis of the top 5 CCs by market capitalization with
traditional asset classes shows the potential of CCs to increase
diversification: As can be seen from \reftab{cor-sas}, correlation
coefficients with none of the traditional assets exceed $0.1$.

\begin{table}
  \centering
  \begin{tabular}{ll}
    \hline \hline
    Name & Asset class \\
    \midrule
    EURO STOXX 50 &	Equity\\
    S\&P100       &	Equity\\
    NIKKEI225	  &	Equity	\\
    FTSE100	  &	Equity	\\
    SSE (Shanghai Stock Exchange) index	&	Equity\\
    MSCI ACWI COMMODITY PRODUCERS 		&	Commodities\\
    GOLD          &	Commodities	\\
    FTSE EPRA/NAREIT DEV REITS& Real Estate\\
    EUR/USD       & Fiat currency\\
    GBP/USD       & Fiat currency\\
    CNY/USD       & Fiat currency\\
    YEN/USD       & Fiat currency\\
    Eurozone 10Y Gov Bonds& Fixed income\\
    UK 10Y Gov Bonds& Fixed income\\
    USA 10Y Treasuries& Fixed income\\
    Japan 10Y Gov Bonds& Fixed income\\
    \hline \hline
  \end{tabular}
  \caption[list-assets]{List of traditional constituents of the investment universe.  Note
    that we term all these asset classes, including alternative assets,
    ``traditional'' in order to contrast them with investments in
    cryptocurrencies (CCs).  We obtain price series for all traditional
    investments from Bloomberg.} 
  \label{tab:list_assets}
\end{table}

%% file: results.tex
In this section we evaluate the out-of-sample performance of the portfolio
allocation strategies in order to address
Questions~\ref{Question4}--\ref{Question6}.  We analyze two dimensions:
First, how does risk-adjusted performance compare across different strategies
and performance measures?  Second, which diversification benefits are
generated by each method?

\subsection{Including CCs in portfolios: performance effects}
The first step of our performance analysis examines how adding CCs to a
portfolio affects efficient frontiers.  In principle, the efficient frontier
is unique, thus identical for all allocation strategies.  However, it depends
on the risk measure (variance or CVaR in this paper), as well as on whether
liquidity constraints are enforced (via LIBRO in this paper) or not.

Our second step then addresses the performance comparison across portfolio
strategies, in terms of cumulative wealth as well as popular risk-adjusted
measures.

\begin{figure}[ht]
  \hspace{-17mm}
  \includegraphics[scale=0.62]{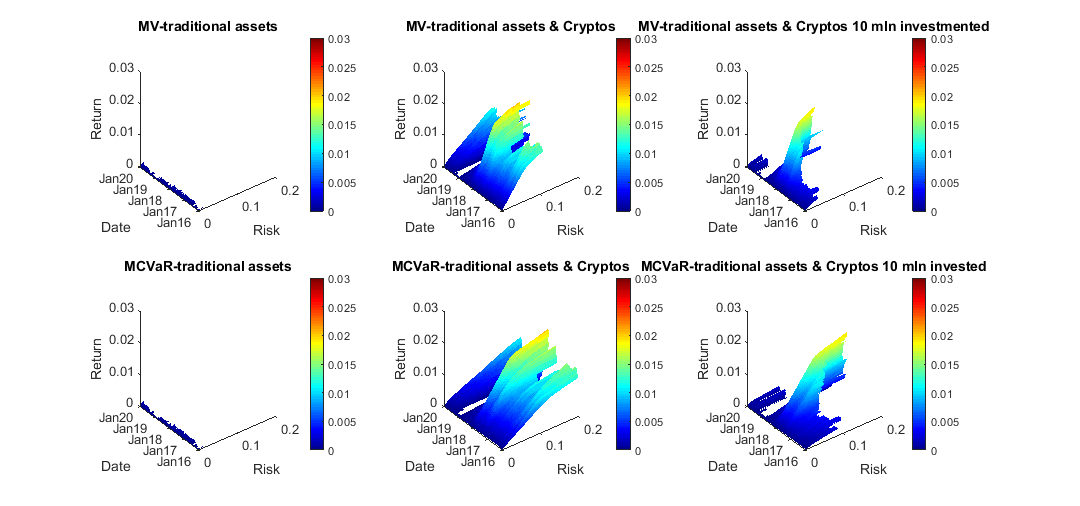}
  \vspace{-7mm}
  \caption[Efficient frontiers]{Efficient frontiers surfaces: the first
    column displays the frontiers for portfolios with only traditional assets
    (including alternative investments, but no cryptocurrencies, CCs) as
    constituents, the second column adds CCs without liquidity constraints,
    and the third column instead adds only CCs up to a liquidity constraint
    (via the LIBRO approach with an investment sum of USD 10 mln).  The top
    row depicts frontiers from mean-variance optimization, the lower one from
    mean-CVaR optimization.  All frontiers are built on a daily basis and
    plotted over the period \periodRange.}
  \label{EffFrontier}
  \quantlet{CCPEfficient_surface}{CCPEfficient\_surface}
\end{figure}

\subsubsection{Efficient frontiers}
\label{eff_front}
\reffig{EffFrontier} plots efficient frontiers for three groups of assets:
only traditional assets, traditional assets \& CCs without liquidity
constraints, and traditional assets \& liquid CCs, up to the constraint
defined via the LIBRO approach with an investment sum of USD 10 mln.  The top
row depicts frontiers from mean-variance optimization, the lower three panels
are based on mean-CVaR-optimal allocations.  All panels show frontiers built
on a daily basis, evolving over time.

For both optimization rules, including CCs leads to a distinct extension of
the frontiers: for low levels of risk, portfolios with CCs give a similar
level of return as without them, but much higher expected returns can be
sought when CCs are included.  The second important observation is that
mean-variance frontiers, in most cases, are shorter than mean-CVaR frontiers
(the same level of returns has lower variance than CVaR), evidence of risk
not being inadequately captured by variance, in line with expectations.  The
LIBRO approach shortens the frontiers especially in the beginning of the
investment period, because it limits the influence of turbulently growing CCs
with low trading volumes.  At the same time, it is visible that starting
roughly in January 2017, the difference between frontiers with (LIBRO) and
without constraints all but vanishes---a change driven by the extreme growth
of trading volumes together with capitalisation of the entire CC market
during that boom period.

The CC market crash in early 2018 is also clearly visible as the frontiers
collapse.  At the trough, series of strongly negative returns amidst high
volatility and evaporating liquidity lead to CCs playing close to no role in
optimal portfolios.  As the market consolidates, in 2019 CCs again pick up
their role in extending the efficient frontier: however, until today the
discrepancy between portfolios with and without concern for liquidity
considerations remains pronounced.  Consequently, the importance of limiting
exposure to illiquid CCs remains high.  Portfolio optimization without
liquidity constraints may promise an attractive performance in theory which
it cannot realize in the market.

\subsubsection{Comparing strategies via performance metrics}
\begin{figure}[ht!]
  \centering
  \includegraphics[scale=1.1]{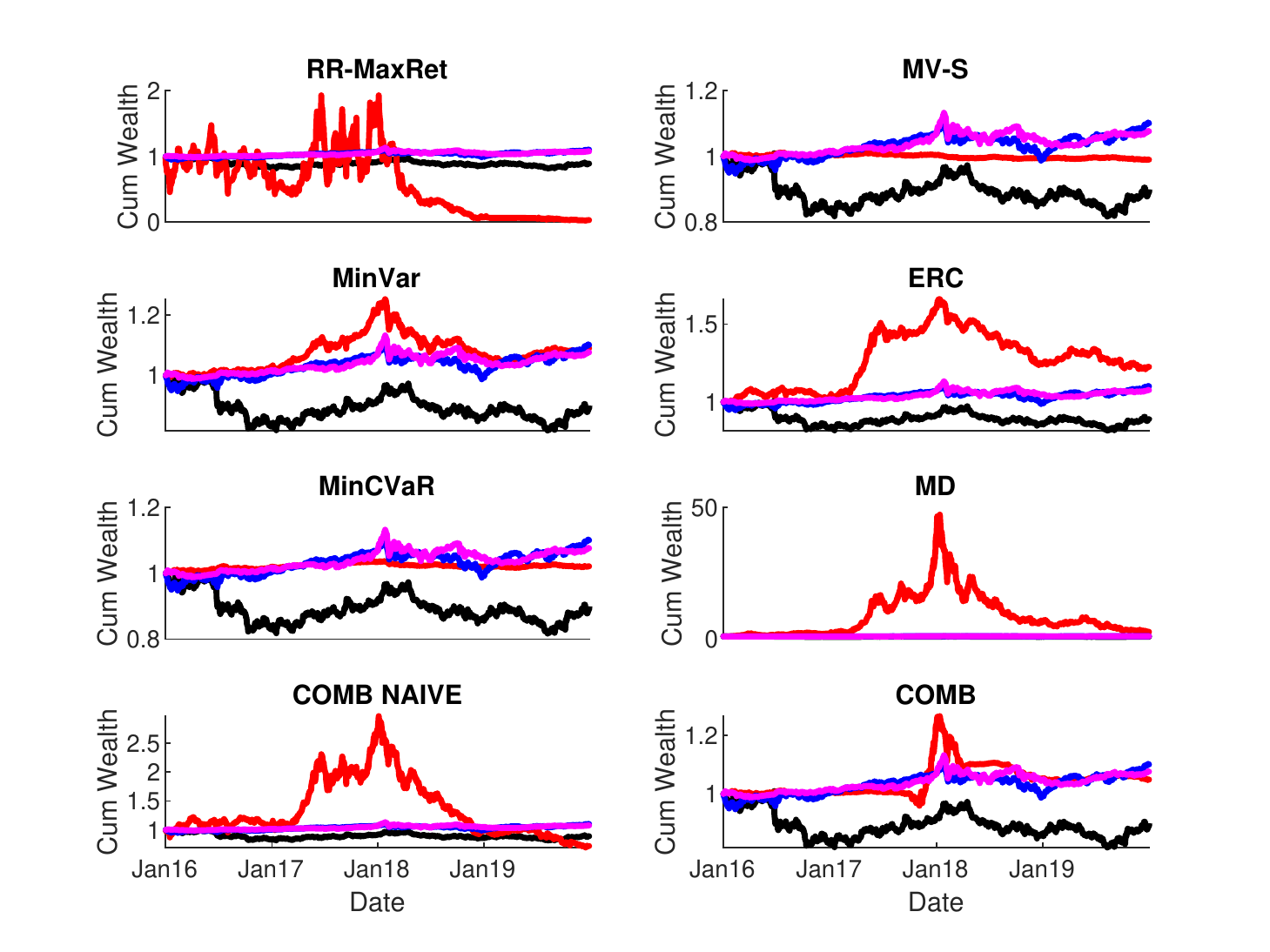}
  \vspace*{-10mm}
  \caption[CW Performance all strats]{\perfinCW\ \emph{without liquidity
      constraints} with monthly rebalancing ($l = 21$) over the period
    \periodRange\ \perflegend.  \trAmeans\ \axesalignment}
  \label{CW_LIBRO_no}
  \quantlet{CCPPerformance}{CCPPerformance}
\end{figure}

\begin{figure}
  \vspace*{-200pt}
  \hspace*{-70pt}
  \includegraphics[scale=.6]{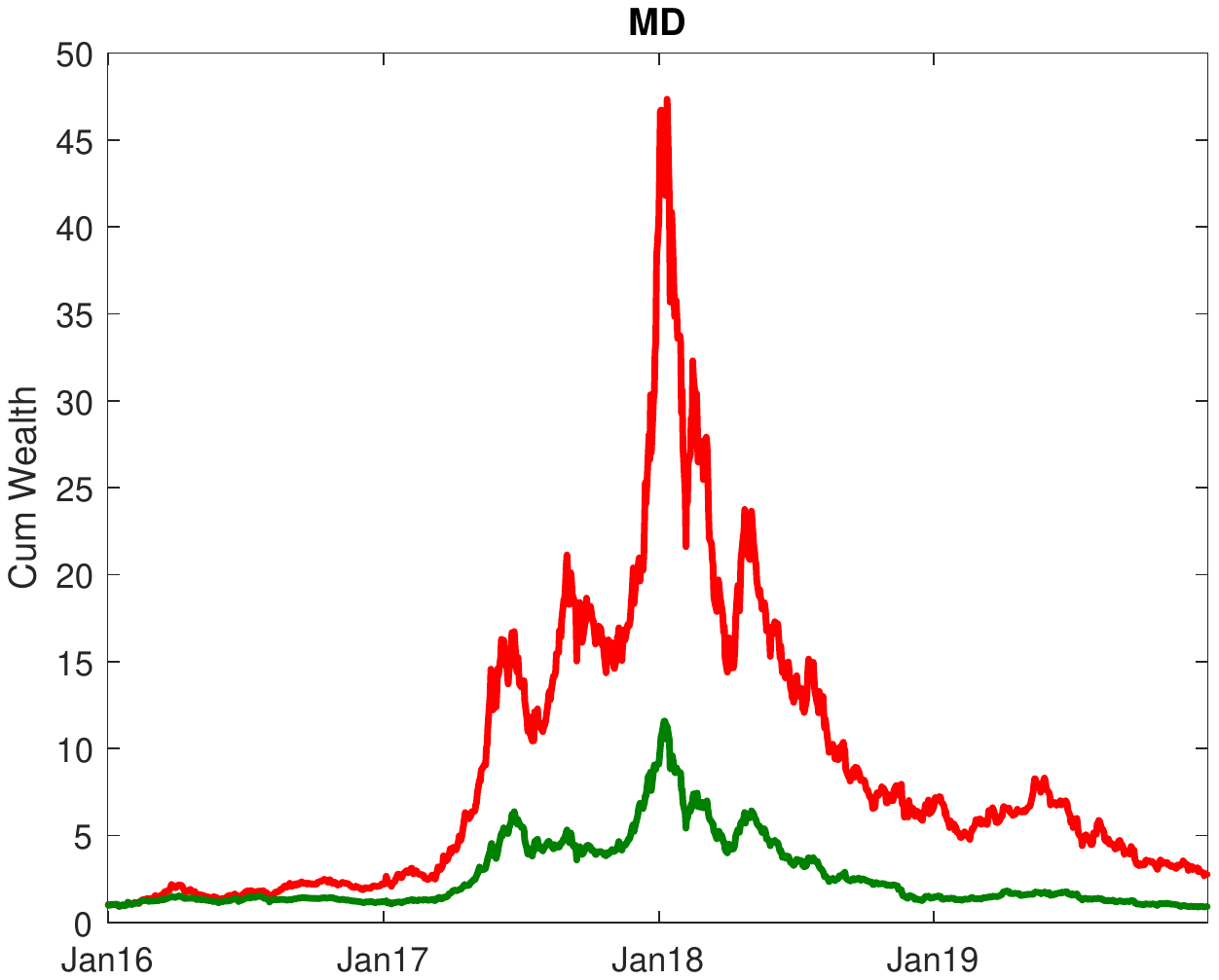}
  \hspace*{-120pt}
  \includegraphics[scale=.6]{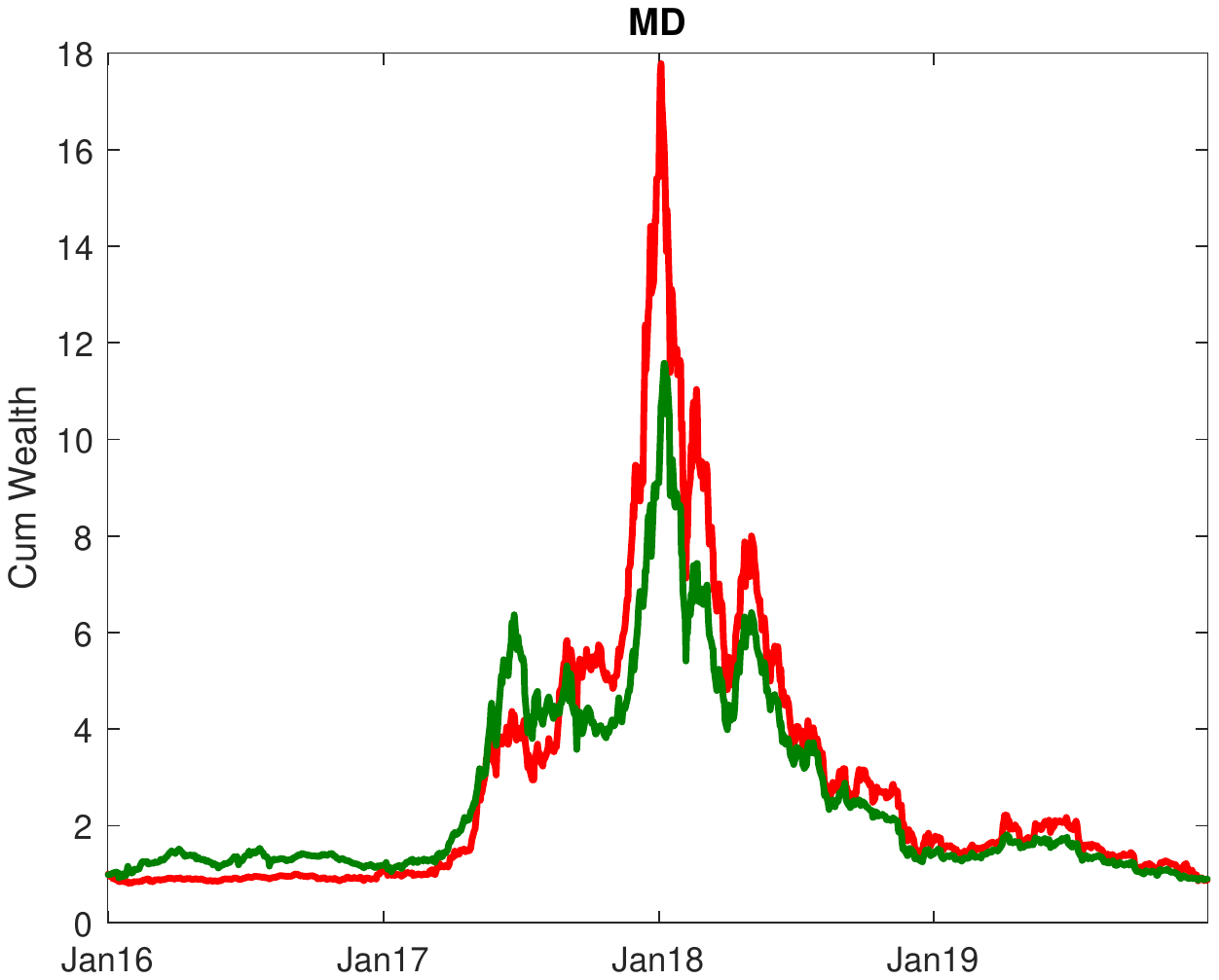}
  \vspace*{-180pt}
  \caption[CUMWEALTH EW with and without LIBRO]{\perfinCW\ of the
    maximum-diversification strategy (MD) without (left panel) and with
    (right panel) liquidity constraints (based on LIBRO at the level of USD
    10 mln), with monthly rebalancing ($l = 21$) over the period
    \periodRange.  For reference, the equally-weighted
    \textcolor{darkgreen2}{EW} strategy is displayed. \axesalignment}
  \label{CUMWEALTH_EW}
\end{figure}

First we examine cumulative wealth, produced by the allocation strategies we
study.  \reffigs{CW_LIBRO_no}{CW_LIBRO_yes} display the dynamics of
cumulative wealth for eight of the strategies considered, with and without
enforcing liquidity constraints, respectively.  As benchmarks we also plot
S\&P100, EW, MV-S and MinVar portfolios built only from traditional
investment constituents (Traditional Assets, ``TrA'').  The EW strategy is
displayed separately in \reffig{CUMWEALTH_EW}, and discussed subsequently.
\reftab{Performance_monthly} summarizes all performance indicators.

The following conclusions can be drawn regarding final Cumulative Wealth (CW)
over the entire period of our study when ignoring liquidity: despite CCs
trading far below their historical peaks at the end of our time span, most
portfolios with CCs generally outperform benchmark portfolios with only
conventional constituents.  However, the discrepancies across strategies are
huge, and the worst-performing strategy RR-MaxRet, which invests always in
the asset with the highest expected return (and thus most often in a CC),
ends up with what can be called a catastrophic result: over the four years of
our study, it loses 97\% of its initial wealth by the end of 2019.
Critically, the strategy did provide stellar results during the boom phase of
2017, exceeding a multiple of 20 times initial wealth at its peak.  Yet
clearly, historical returns were no long-term predictor of expected returns
for the best-performing CCs, and the lack of diversification hurt this
strategy badly.

On the other end of the spectrum, the highest result is achieved by MD, with
an accumulated final wealth of 275\%.  This amounts to an annualized rate of
return of just below 30\% over a 4-year period in which the S\&P100 lost
10\%.  Critically, this result is also achieved by investing in small CCs
(and therefore also follows the boom-and-bust cycle to a comparable degree):
the difference is driven by the very strong diversification the MD strategy
pursues by design.  It is therefore not surprising that ERC turns out the
second-best strategy, with a +22\% return over the period.  Its construction
successfully limits its exposure to the extremes during 2017/18 to about an
order of magnitude lower than MD.

Regarding the combined strategies, the na\"ive version is strongly
susceptible to RR-MaxRet, while the bootstrapped version performs quite well.

Finally, the model-free EW strategy with CCs underperforms with a final loss
of 13\%, while equal weighting across only traditional assets achieves the
best performance among the benchmark strategies.  However,
\reffig{CUMWEALTH_EW} shows how EW performance exhibits high variation over
the time span, similar in nature to MaxRet and MD.  The figure displays MD
and EW separately, to elucidate two important points: first, how
disproportionately the performance of small coins exceeded the gains of
established CCs in the 2017 price explosion; second, how seriously calculated
results of portfolio allocation rules can diverge from returns achievable by
investors if lack of liquidity is not taken into account.

Generally, LIBRO portfolios have mixed results in terms of cumulative wealth.
Most importantly, MD underperforms when enforcing LIBRO constraints.  Of
course, this implies that the high performance of unconstrained optimization
can only be reached for very small investment sums.  For larger portfolios,
when the liquidity constraints turn binding, performance need not necessarily
suffer.  By limiting the exposure to individual (and thus also small) coins,
some strategies, including RR-MaxRet, are positively affected by LIBRO.  When
ignoring the liquidity risk, this strategy retained 3\% of its initial value;
with LIBRO it retains 59.1\%.  Also the combined strategy COMB, which
provides a positive performance without LIBRO, further improves by 8.6\% when
protecting the portfolio from liquidity risk.


\begin{figure}[ht!]
  \centering
  \includegraphics[scale=1.1]{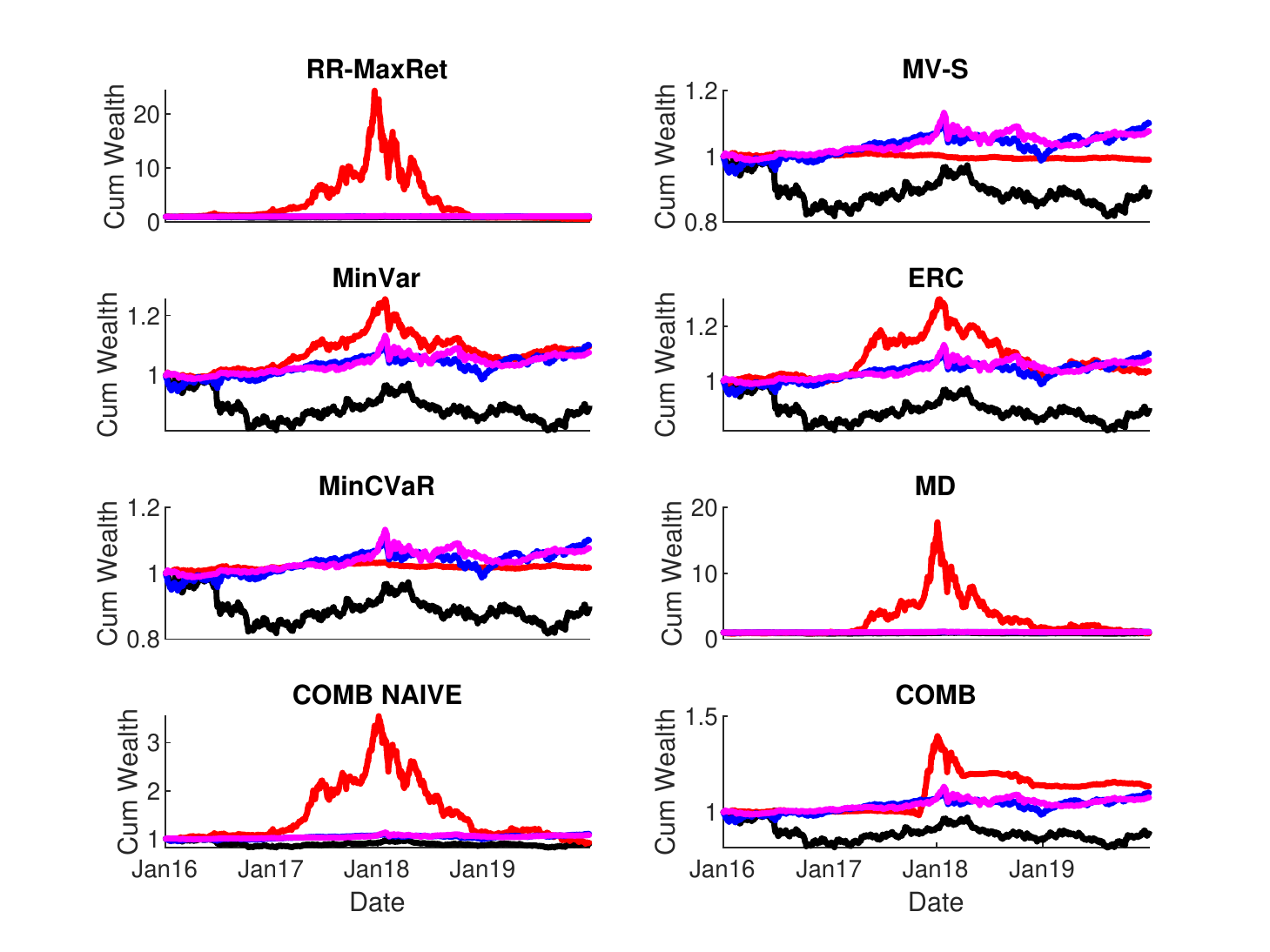}
  \vspace*{-10mm}
  \caption[CW Performance all strats LIBRO]{\perfinCW\ \emph{with liquidity
      constraints} (based on LIBRO at the level of USD 10 mln) and monthly
    rebalancing ($l = 21$) over the period \periodRange\ \perflegend.
    \trAmeans\ \axesalignment}
  \label{CW_LIBRO_yes}
  \quantlet{CCPPerformance}{CCPPerformance}
\end{figure}

\tab{Performance_monthly}{\perfmeasures, with monthly rebalancing ($l =
  21$). \measuresAre\ \trAmeans\ Strategies are detailed in
  \reftab{list_strategies}.  \superiorperfred\ 
}

\begin{table}
  \centering
  \input{tables/Pvalue_monthly_liquidity_constraint_no}
  \caption[Pvalue-monthly-liquidity-constraint-no]{Tests for difference
    between the Sharpe ratio SR (lower triangle) and the certainty equivalent
    (CEQ, upper triangle) of all strategies with respect to each other:
    color-coded $p$-values with significance at the
    \textcolor[HTML]{B80000}{\bf 0.01}, \textcolor[HTML]{FF0E0E}{\bf 0.05}
    and \textcolor[HTML]{FF6363}{\bf 0.1} level (without liquidity
    constraints).}
  \quantlet{CCPTests}{CCPTests}
  \label{tab:Pvalue_monthly_liquidity_constraint_no}
\end{table}
 
\begin{figure}[ht!]
  \includegraphics[scale = 0.9]{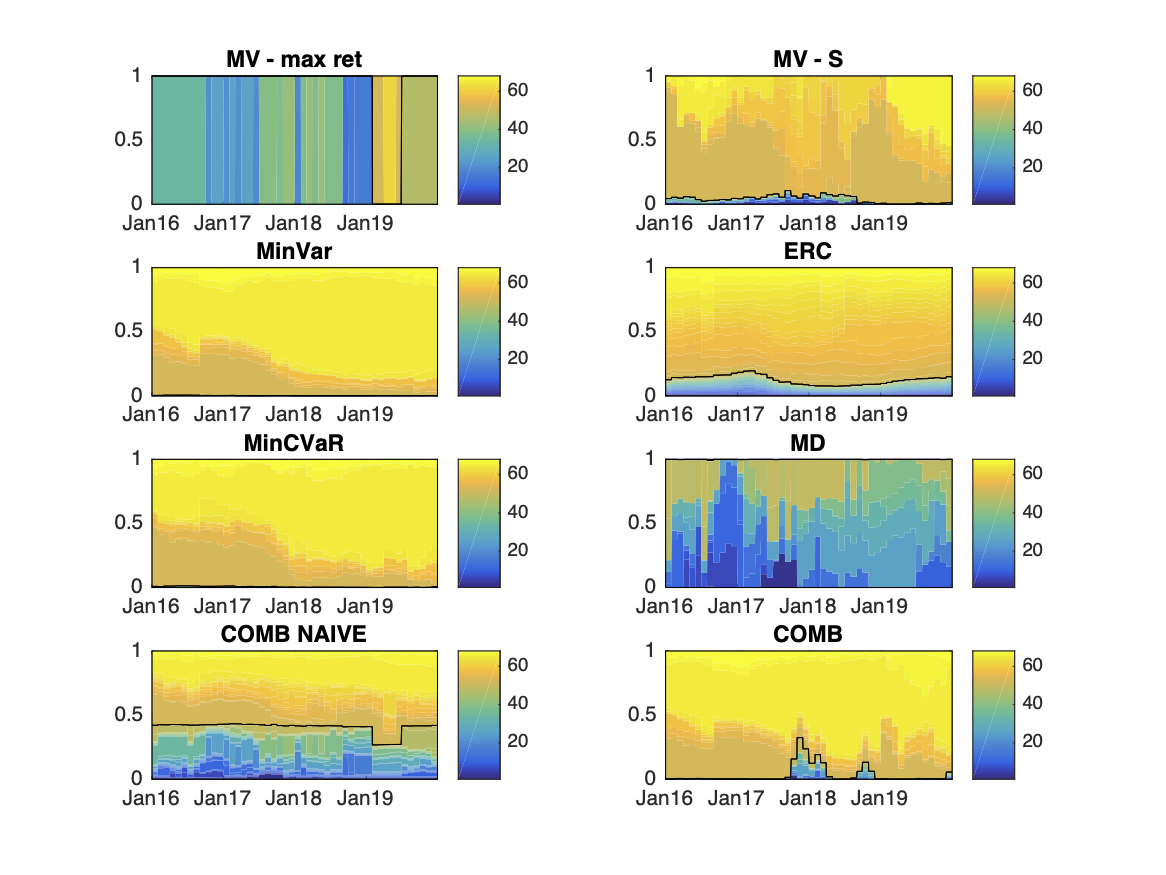}
  \vspace{-10mm}
  \caption[weights-dist no-libro]{Evolution of the \emph{portfolio
      composition} (i.e., relative weights) of all allocation strategies
    (without liquidity constraints) with monthly rebalancing over the period
    \periodRange: \blacklineseparates.}
  \quantlet{CCPWeights}{CCPWeights}
  \label{Weights_LIBRO_no}
\end{figure}
\begin{figure}[ht!]
  \includegraphics[scale = 0.9]{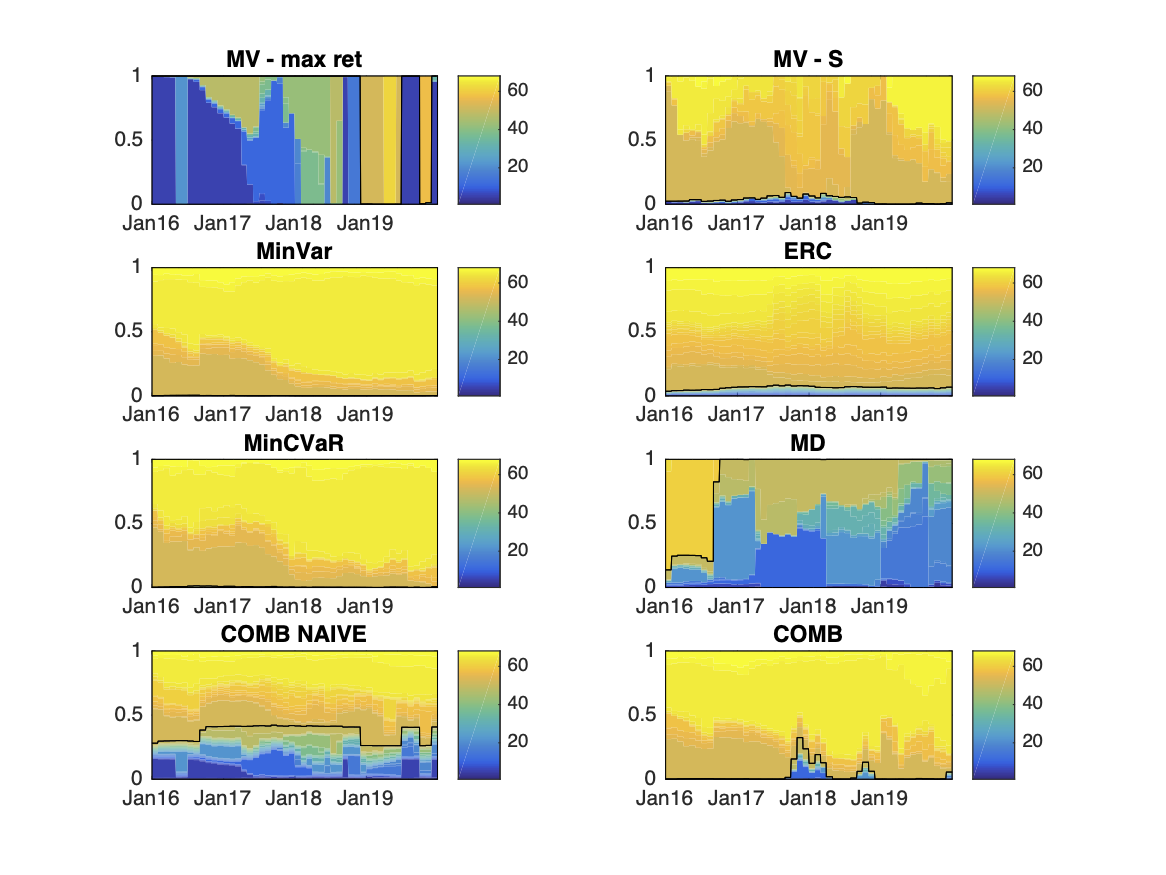}
  \vspace{-7mm}
  \caption[weights-dist with-libro]{Evolution of the \emph{portfolio
      composition} (i.e., relative weights) of all allocation strategies
    \withlibrolimits\ with monthly rebalancing over the period \periodRange:
    \blacklineseparates.}
  \label{Weights_LIBRO_yes}
  \quantlet{CCPWeights}{CCPWeights}
\end{figure}

Next, we analyse risk-adjusted performance for all portfolios.  While MD
demonstrates superior absolute performance, ERC dominates in terms of
risk-adjusted performance, in particular in terms of its (adjusted) Sharpe
Ratio of $0.033$.  Importantly, turnover is much lower at $4.2$
(unconstrained, constrained: $9.8$), slightly below that of EW and above
MV-S.  Turnover and target turnover per strategy are reported in the last
four columns in \reftab{Performance_monthly}, which show that the only
strategy with appreciably lower trading is RR-MaxRet at $0.69$ (constrained:
$0.73$), with the above-mentioned harsh result.  This is expected, given that
the strategy is by construction the most concentrated one, consisting of the
one asset with the highest return (see also
\reffigs{Weights_LIBRO_no}{Weights_LIBRO_yes}).

It is interesting to note that for the strategies with strong
diversification, in particular MD and ERC, but also MV-S, enforcing the LIBRO
constraints leads to \emph{higher} turnover.  This is of concern to
investors, as it prompts higher transaction costs.  At first sight this
observation appears counterintuitive, as restricted weights could be expected
to reduce trading needs (due to positions partially remaining at their
binding limits).  The puzzle is explained by the last two columns, reporting
target turnover: clearly, changes \emph{in target weights} are mitigated via
the liquidity constraints, corresponding to intuition.  At the same time, it
is exactly small and illiquid CCs which exhibit the largest volatility, and
thus prompt \emph{larger} trades when at the next rebalancing date positions
are brought back to target weights.  Enforcing LIBRO constraints leads to
positions in more (and prone to be smaller) CCs, triggering larger
rebalancing needs in terms of portfolio turnover.

Finally, \reftab{Pvalue_monthly_liquidity_constraint_no} reports when the
differences between strategies in terms of CEQ or SR are significant, based
on tests described in \refapp{SRdiff}.  Although MD, COMB, and in particular
MV-S have SR and CEQ higher than the EW strategy, tests do not support
significance of this difference.  In constrast, the ERC portfolio exhibits a
higher SR and this difference is significant.  The comparison of
risk-adjusted metrics for MinVar and MinCVaR reveals that they differ
significantly from each other---testament to the strong deviation of CC
returns from the normal distribution.  MinCVaR also differs significantly
from the diversifying strategies MD and ERC.

As a robustness check, we also conduct all analyses for weekly and daily
rebalancing of portfolios.  Results are provided in \refapp{PerformanceApp},
generally confirming the conclusions so far, and show that the qualitative
results are robust with regard to the rebalancing frequency.

\subsection{Including CCs in portfolios: diversification effects}
We separately analyse diversification characteristics of the allocation rules
for two reasons: On the one hand, CCs are known from the literature for their
diversifying properties; on the other hand, the most diversifying strategies
MD and ERC performed best.  First, we examine the composition of the optimal
portfolios over time.  Second, we run mean-variance spanning tests in order
to establish if CCs are a valuable addition to broadly diversified portfolios
of traditional assets.  Third, we analyse diversification across the
portfolio strategies by means of dedicated diversification measures.
 
\subsubsection{Portfolio composition}
\reffigs{Weights_LIBRO_no}{Weights_LIBRO_yes} plot the evolution of portfolio
constituents across time, without and with liquidity constraints,
respectively.  At each date on the abscissa, the simplex of weights is
color-coded vertically, with traditional assets on the light end of the
spectrum and CCs towards the dark end; a black lines indicates the boundary
between the two groups.  We can see wide variation in the extent to which the
strategies rely on CCs: MaxRet and MD are prone to invest heavily in CCs,
while risk-oriented strategies like MinVar and MinCVaR hardly include any.
The risk-return-oriented strategy MV-S employs CCs conservatively, yet it
does reach at times noteworthy allocations even against the background of
such a well-diversified portfolio of traditional assets.  The share of CCs is
lower in the last 2 years of the time period, but does not drop to zero.

Most importantly, the figures point out how the LIBRO approach, as expected,
significantly affects portfolio weights; the most visible difference arises
for models with a high share of CCs, namely MD and RR-MaxRet, but also ERC,
where it mitigates the exposure particularly in the first half of the
investment period.

The weights distribution of the COMB portfolio undergoes quite pronounced
changes over the investment period: from high concentration of traditional
assets to high concentration of CCs, and back---confirming that no individual
model outperforms its competitors permanently.

To shed more light on how these weights affect the performance of each
strategy's portfolio, we also compare the risk structures for all strategies
in \reffigs{Risk_LIBRO_no}{Risk_LIBRO_yes}.  After all, the volatility
structure of CCs leads to disproportionate risk contributions relative to
their capital weights: traditional assets affect changes in portfolio values
to a visibly lower degree.

\subsubsection{Mean-variance spanning}
In order to investigate the impressions from the efficient-frontier plots in
\refsec{eff_front}, we conduct two mean-variance spanning tests on each of
the 52 CCs: first, the corrected test of \textcite[HK,][]{huberman1987mean},
second the step-down test by \textcite{kan2012tests}.

\tab{SpanTest}{Spanning Tests for individual cryptocurrencies with respect to
  the efficient frontier constructed from all traditional investment assets,
  including alternative assets (see \reftab{list_assets} for a complete list;
  $p$-value in parentheses).  F-Test refers to the corrected test of
  \textcite{huberman1987mean}, $F_1$ and $F_2$ to step-down tests by
  \textcite{kan2012tests}, testing for spanning of tangency portfolios and
  for global minimum portfolios, respectively.  Only CCs for which at least
  one test rejects spanning at the 10\% level are reported.}

\reftab{SpanTest} lists only CCs with at least one test rejecting the
hypothesis that traditional assets span the frontier at the 10\% level.
Recall that our definition of traditional assets includes a broad set of
alternative investments, all but CCs.  The corrected HK test rejects spanning
for 3 CCs.  In contrast, the step-down test provides information on the
source for spanning rejection: $F_1$ tests for spanning of tangency
portfolios, whereas $F_2$ tests spanning for global minimum portfolios.  From
\reftab{SpanTest}, we see that the $F_1$ test rejects spanning for only 2
CCs, pointing out that tangency portfolios which include CCs are
significantly different from the benchmark tangency portfolio, but also that
the inclusion of the two years 2018--19 has dramatically reduced that number
from previously 27 CCs, which included Bitcoin (BTC), Ripple (XRP), Dash
(DASH) and Litecoin (LTC).  $F_2$ rejects spanning for 5 CCs for the entire
time period, still including one of the coins with the highest market
capitalisation, XRP.  Thus, we conclude there still exists evidence that a
MV-S portfolio can be improved by 7 out of 52 CCs, but that the integration
of CC with financial markets has progressed markedly.  Anecdotal evidence in
line with this finding comes from the recent outbreak of the corona-virus
pandemic, when initially CC markets moved for the first time with strong
positive correlation together with financial markets, driven by institutional
investors rebalancing in favor of cash holdings, before CCs resumed their
diversifying role in subsequent weeks.

Also, but there is little evidence that a MinVar portfolio can be improved.
This result is supported by the dynamics of the portfolios' composition
presented in \reffigs{Weights_LIBRO_no}{Weights_LIBRO_yes} for unconstrained
and LIBRO portfolios, respectively: MinVar portfolios in both cases are
constructed entirely from traditional assets, whereas MV-S portfolios have a
(varying) CC component throughout the whole investment period.

In sum, the results imply that investors should consider a broader selection
of CCs (see Question~\ref{Question2}), not only BTC.  However, only a small
fraction of CCs continue to improve the efficient frontier.

\subsubsection{Diversification metrics across portfolio strategies}
\reftab{Diversification_monthly} reports results on our three chosen
diversification metrics (detailed in \refsec{divers_meas}).  As expected, for
the RR-MaxRet strategy there are no diversification benefits---by definition
it consists of only one asset at a time (unless LIBRO forces it into more
than one asset).  The range of values across diversification metrics
emphasizes that diversification has different aspects and its quantification
depends on the definition used.  Consequently, different measures do not
always provide identical conclusions about the diversification effects of CCs
in portfolios.  For instance, in terms of a $DR^2$ of $13.73$
($13.44$),\footnote{Here and henceforth we provide the values of the
  performance metric for LIBRO portfolios in parentheses.} MinCVaR is
characterized as most diversified strategy.  Slightly lower measures pertain
for MinVar and ERC portfolios with $12.02$ ($11.72$) and $8.71$ ($9.41$),
respectively.  The MD portfolio is a special case regarding this type of
diversification, with $DR^2$ of $2.64$ ($2.07$) and at the same time a PDI of
$21.06$ ($21.01$).  Clearly PDI highest for the MD portfolio because of its
objective function, maximizing diversification via the number of independent
sources of variation in the portfolio.

\tab{Diversification_monthly}{\diversifmeasures, with monthly rebalancing
  ($l = 21$).  \diversifmeasuresexplained\ \superiorperfred}
 
The ERC portfolio is characterized by the highest Effective $N$ of $17.16$
($13.61$) by a large margin, also a typical result \citep[see,
e.g.,][]{clarke2013risk} due to its nature: it includes all assets by
definition.  Apart from MaxRet, the lowest Effective $N$ of $2.68$ ($2.69$)
arises for the MinVar portfolio, containing only traditional assets, showing
that fewer than $3$ equally-weighted stocks would provide the same
diversification by this measure.  All other individual strategies also
exhibit Effective $N$ ranging between 3 and 4.  One more remarkable result
concerns the combined portfolios' concentration: While COMB's Effective $N$
lies in the range of individual strategies, COMB~Na\"ive exceeds $10$ both in
constrained and unconstrained portfolios.  In terms of $DR^2$, the combined
strategies rank inversely, reaching $3.43$ ($3.62$) for COMB~Na\"ive and
$10.0$ ($9.44$) for COMB; their PDIs are similar to those of the other
risk-oriented portfolios MinVar, MinCVaR and ERC.

Note that with the exception of ERC liquidity constraints do not strongly
affect the diversification features of portfolios: all metrics display only
minor changes.  This result is due to the fact that LIBRO generally lowers
the weight of constituents, but does not completely exclude them.

We particularly highlight the difference of diversification of the MV-S
portfolios with and without CCs: diversification measured by $DR^2$ increases
through the inclusion of CCs from $5.39$ ($5.39$) to $8.03$ ($7.48$), and PDI
scales up even more strongly, from $4.92$ ($4.92$) to $20.62$ ($20.62$).
Thus, the inclusion of CCs remarkably improves portfolio diversification,
especially in terms of the distribution of principal portfolio variances.

\subsection{Interpretation of the results}
In this section we relate our empirical results to the
Questions~\ref{Question4}--\ref{Question6}, along which the contribution of
this paper is structured.

\newcounter{questioncounter}
\setcounter{questioncounter}{0}
\newcommand{\repeatQu}[1]{\emph{Question \stepcounter{questioncounter}
    \thequestioncounter: #1}}

\repeatQu{\questionFour}

As the efficient frontiers in \reffig{EffFrontier} clearly show, the main
benefit of CCs accrue to investors who make use of the high-risk/high-return
character of their returns; investors with low risk tolerance benefit least.
While it is not surprising that CCs constitute risky investments,
\reffig{Weights_LIBRO_no} shows how the risk-oriented strategies (minimizing
variance or CVaR) consist almost entirely of traditional assets, so CCs have
no influence on them; at least a risk-return orientation is necessary for CCs
to play a noteworthy and permanent role in portfolios.  At the other end of
the specturm, the extremely CC-affine MaxRet strategy, despite stellar
performance during the boom phase of 2017, was all but wiped out by the end
of our period (ultimately retaining only 3\% of its initial value).

The model-free EW strategy is a special case: its performance in the middle
of our time period was extraordinary, and so was its collapse when the 2017
price rally in CCs disintegrated.  As with MaxRet, both parts are driven by
the high weight of small CCs---these were precisely the ones that gained
disproportionately in value during the price rally, and subsequently suffered
the severest.  Therefore, over our complete time span the EW portfolio in
fact lost $12.3\%$ in value, whereas other types of investors ended up with
gains.

By far the best performance was achieved by investors who target strong (or
even maximal) diversification.  These strategies, ERC and MD, lead to
sizeable exposures to a broader cross-section of CCs, while they limit the
risks the EW strategy incurs.

The general conclusion is that the utility from adding CCs to a portfolio
strongly depends on the investor's objective.  In particular investors
targeting a well-diversified portfolio while willing to bear some risk are
advised to consider CCs for their investments.

\repeatQu{\questionOne}

The rebalancing frequency (whether portfolio positions are traded daily,
weekly, or monthly to react to market developments by updating estimates and
to revert positions to target weights) does influence the performance of
investors' portfolios.  For instance, over our study period cumulative wealth
for the MV-S strategy grows by 5\% when readjusting the portfolio on a daily
basis, by 12\% with weekly, and 9\% with monthly position changes.  These
difference become more pronounced when transaction costs are deducted, as
turnover is naturally higher at a higher rebalancing frequencies.  For the
RR-MaxRet strategey, the loss attenuates with weekly and exacerbates with
daily reallocations.

However, the overall picture does not change across rebalancing frequencies:
Even with a more frequent reallocations, still diversification-seeking
investors (ERC and MD) significantly outperform the other investment
strategies.  Therefore, our general conclusions about the effect of adding
CCs into investment portfolios do not change qualitatively between daily
traders, weekly rebalancing and monthly reallocation (retail investors).

\repeatQu{\questionTwo}

Most importantly, our findings clearly indicate that diversification also
across CCs is beneficial.  At the same time, investors could diversify too
much.  As \reftab{Diversification_monthly} shows, the MD strategy, which had
the highest return, showcases an Effective $N$ of only $3.26$.  ERC has much
higher Effective $N$ of $17.16$, still it features considerably lower final
cumulative wealth, at least in unconstrained optimization.  Judged by PDI, MD
is the most successful strategy, which of course is driven by the fact that
the target-weight allocation of MD is derived precisely by maximizing PDI.
However, this also indicates that including as many assets in the portfolio
as possible is not necessary to adequately represent the covariance matrix,
and not beneficial in terms of cumulative wealth.

\reffigs{Weights_LIBRO_no}{Weights_LIBRO_yes} caution the interpretation of
MD dominating in terms of accumulated returns.  Both Figures show that MD
includes a broad range of CCs, whereas MinVar and MinCVaR---both with
comparable Effective N and PDI---almost entirely exclude them, giving weight
only to traditional assets.  In this sense we do find evidence that CCs can
substitute for traditional assets in portfolio optimization.

Regarding the ERC strategy, while it reaches optimal diversification for the
alternative metric of Effective $N$, it provides sizable gains in cumulative
wealth and at the same adequately diversifies the portfolio.
\reffigs{Weights_LIBRO_no}{Weights_LIBRO_yes} indicate that CCs and
traditional assets are mixed in the portfolio, while the PDI is close to the
one of MD and $DR^2$ only second to the pure risk-oriented strategies MinVar
and MinCVaR.

Therefore, including CCs to diversify the portfolio is beneficial to achieve
high target returns, and balancing traditional assets and CCs is advisable.

\repeatQu{\questionThree}

Even though CCs are highly volatile, the past pricing series are informative
for portfolio allocation.  As such, quantitative methodologies for portfolio
allocation are applicable and one is not restricted to non-quantitative or
model-free investment schemes.  The EW strategy, which we discussed in the
answer to Question~1, can exhibit phases of extraordinary returns, but does
not manage risk well.  At the other end of the spectrum, however, strategies
exclusively targeted at lowering risk at all cost do not benefit from CCs.
This is of course unsurprising, since lower risk must go at the expense of
lesser expected return, most clearly visible already in the efficient
frontiers in \reffig{EffFrontier}.

\repeatQu{\questionFive}

This question is addressed by LIBRO: The fact that the bounds on CC weights
by LIBRO \citep{trimborn_investing_2017} turn out to bind indicates that
several CCs are not sufficiently liquid for investors with deeper pockets.
Still, the approach allows the inclusion of illiquid CCs up to restricted
amounts.  This has the positive effect that investors can still perform
diversification strategies to quite some degree---strategies that rank among
the most profitable.  However, the impressive results by strategies with
broad CC exposure turn out not to be very scalable.  For instance, the MD
strategy shows excellent performance without liqudity constraints ($+175\%$),
yet the application of LIBRO pushes final CW below initial wealth ($-14\%$).

Clearly, the performance of unconstrained MD profits from unreasonably high
weights on small and illiquid CCs.  \reftab{Diversification_monthly}
illustrates this in terms of Effective $N$ and PDI: MD reaches a very low
Effective $N$ of $3.99$, although it includes only CCs, compare the weight
composition in \reffigs{Weights_LIBRO_no}{Weights_LIBRO_yes}.  PDI is clearly
higher than for other strategies, in line with the objective of MD, and the
PDI only shrinks marginally when incorporating LIBRO, whereas Effective $N$
drops by about 1.  This implies that the strategy focuses disproportionately
on (a) singular CC(s), driving the high returns, which cannot be traded
sufficiently for a portfolio of USD 10 mln.  At the other end of the
spectrum, the mininum-risk strategies focus on traditional assets with high
trading volume, therefore they are little affected by LIBRO.

\repeatQu{\questionSix}

Regarding CC return properties, a lot of prior research exists (see also our
literature review in \refsec{lit_review}).  We confirm that the patterns of
generally high means, high volatilities, excess kurtosis, and low
correlations with traditional assets are also present in our sample (see the
descriptive statistics in \refapp{DescrStatApp}).  Our contribution addresses
the effect of including CCs in already broadly diversified portfolios: Beyond
what we have established in the answers to the preceding five questions, our
central finding is that the key conclusion of prior studies---that CCs are
valuable additions to the investment universe---holds true, but it is
critical to mind the limits of quantitative results derived from simplified
frameworks.  While diversification strategies prove most promising, including
only the top CCs foregoes diversification potential.  Most importantly,
returns of broad CC portfolios that are calculated without accounting for
liquidity remain virtual: they cannot be realized by professional investors.

Finally, for certain types of investors, namely those highly risk averse, the
benefits can prove too risky to pursue.

%
%


%% file: tables/Pvalue_monthly_liquidity_constraint_no.tex
\begin{tabular}{l@{ }l|p{5pt}!{\color{white}\VRule[2pt]}p{5pt}!{\color{white}\VRule[2pt]}p{5pt}!{\color{white}\VRule[2pt]}p{5pt}!{\color{white}\VRule[2pt]}p{5pt}!{\color{white}\VRule[2pt]}p{5pt}!{\color{white}\VRule[2pt]}p{5pt}!{\color{white}\VRule[2pt]}p{5pt}!{\color{white}\VRule[2pt]}p{5pt}!{\color{white}\VRule[2pt]}p{5pt}!{\color{white}\VRule[2pt]}p{5pt}!{\color{white}\VRule[2pt]}p{5pt}!{\color{white}\VRule[2pt]}p{5pt}!{\color{white}\VRule[2pt]}p{5pt}}

\multicolumn{2}{c}{Allocation strategy} & 1& 2& 3& 4& 5& 6& 7& 8& 9& 10 & 11 \\
\toprule
1 & S\&P100&&&&\cellcolor[HTML]{B80000}&&&&&&&\\\arrayrulecolor{white}\hline\hline
2 & EW-TrA&&&&\cellcolor[HTML]{B80000}&&&&&&\\\arrayrulecolor{white}\hline\hline
3 & EW&&&&\cellcolor[HTML]{B80000}&&&\cellcolor[HTML]{FF6363}&&&&\\\arrayrulecolor{white}\hline\hline
4 & RR Max Ret&&&&&\cellcolor[HTML]{B80000}&\cellcolor[HTML]{B80000}&\cellcolor[HTML]{B80000}&\cellcolor[HTML]{B80000}&\cellcolor[HTML]{B80000}&\cellcolor[HTML]{B80000}&\cellcolor[HTML]{B80000} \\\arrayrulecolor{white}\hline\hline
5 & MV-S&&&&&&&&&&&\\\arrayrulecolor{white}\hline\hline
6 & MinVar&&&&&&&&\cellcolor[HTML]{FF0E0E}&&&\\\arrayrulecolor{white}\hline\hline
7 & ERC&&&\cellcolor[HTML]{B80000}&\cellcolor[HTML]{FF6363}&&&&&&\cellcolor[HTML]{FF6363}&\\\arrayrulecolor{white}\hline\hline
8 & MinCVaR&&&&\cellcolor[HTML]{FF6363}&&\cellcolor[HTML]{FF0E0E}&\cellcolor[HTML]{FF0E0E}&&\cellcolor[HTML]{FF0E0E}&\cellcolor[HTML]{FF0E0E}&\cellcolor[HTML]{FF6363}\\\arrayrulecolor{white}\hline\hline
9 & MD&&&&\cellcolor[HTML]{FF6363}&&&&&&&\\\arrayrulecolor{white}\hline\hline
10 & COMB NA\"IVE&&&&&&&\cellcolor[HTML]{B80000}&&&&\\\arrayrulecolor{white}\hline\hline
11 & COMB&&&&\cellcolor[HTML]{B80000}&&&&&&&\\\arrayrulecolor{white}\hline\hline\\
\bottomrule
\end{tabular}

%% file: conclusion.tex
This study investigates cryptocurrencies as new investment assets available
to portfolio management. We investigate the utility gains for different types of investors when they consider cryptocurrencies as an addition to a well diversified portfolio of traditional assets. We consider risk-averse, return-seeking as well as diversification-preferring investors who trade along different allocation frequencies, namely daily, weekly or monthly. To conduct this study, we analyze the performance of commonly used asset-allocation models based on historical prices and trading
volumes of \XsecN\ cryptocurrencies, combined with 16 traditional assets.
The rules-based investment methods cover a broad spectrum of investor
objectives, from the classical Markowitz optimization to recent strategies
aiming to maximize portfolio diversification.  Along with individual
portfolio allocation strategies, we also include combined strategies from
model averaging.  The performance of portfolios is evaluated with a range of
different measures, including cumulative wealth, risk-adjusted performance
and diversification effects produced by portfolios.
  
We find that due to the volatility structure of cryptocurrencies, the
application of traditional risk-based portfolios strategies, such as
equal-risk contribution, minimum-variance and minimum-CVaR, does not boost
the performance of investments significantly.  In contrast, approaches such
as the maximum-return strategy (or strategies with high target returns), and
also the maximum-diversification portfolio reach higher expected returns via
higher or broader cryptocurrency exposure for investors.  As for
diversification benefits, we demonstrate an effect beyond well-diversified,
global portfolios of conventional assets without CCs.  We also document how
various rules have different effects on portfolio diversification, depending
on the concept of diversification and the chosen measure of its
quantification.
  
Furthermore, following the idea of model averaging and diversification across
models, we show that both naive and bootstrap-based combined portfolios
exhibit robust high risk-adjusted returns.  Portfolios with model-averaged
weights achieve significantly higher performance than purely risk-oriented
strategies and not significantly lower than the best performing strategies.
  
We also show how different rebalancing frequencies affect performance, as
well as how constraints mitigating liquidity risks of cryptocurrencies
(LIBRO) can significantly affect the outcome of strategies that rely on a
larger cross-section of CCs.  The results remain coherent across all
frameworks.  Further extensions can be made along three main lines: first,
more involved estimators of expected returns and the covariance matrix could
be employed; second, more performance measures could be used to evaluate the
investment strategies' results; and third, additional portfolio-allocation
strategies could be included in the comparison.  In particular, factor-based
APT (arbitrage price theory) models would constitute the complementary
approach to statistical-optimisation techniques studied in this paper.

%

%% file: appendix.tex
\subsection{Test for difference of SR or CEQ between two strategies}
\label{SRdiff}
\input{App_Wolf_Ledoit}

\clearpage

\subsection{Descriptive statistics of portfolio components}
\label{DescrStatApp}
\input{App_Descr_stat.tex}

\clearpage

\subsection{Dynamics of risk contributions for portfolio strategies}
\label{RiskContrib}
\input{App_RiskContrib.tex}

\clearpage


\newpage
\subsection{Results  for daily and weekly  rebalanced portfolios}
\label{PerformanceApp}
\input{App_Performance}

%% file: App_Wolf_Ledoit.tex
We employ the test by \textcite{ledoit2008robust}.  Let
$\nu = ( \mu_i, \mu_j, \sigma_i, \sigma_j)$ denote the vector of the moments
of two strategies $i$ and $j$.

Then we can test for a difference of the strategies' CEQs or SRs via the test
statistics defined as the differences of those measures,
\begin{equation}
  f_{CEQ}(\nu) = \mu_i-\frac{\gamma}{2}\sigma^2_i-\mu_j+\frac{\gamma}{2}\sigma^2_j,
\end{equation}
or
\begin{equation}
  f_{SR}(\nu) = \frac{\mu_i}{\sigma_i}-\frac{\mu_j}{\sigma_j},
\end{equation}
respectively.

Applying the delta method yields that if
$\sqrt{T-M} (\hat{\nu}- \nu)\xrightarrow{\text{d}}N(0, \Psi)$, then
\begin{equation}
  \sqrt{T-M} (\hat{f}-f)\xrightarrow{\text{d}}N(0, \nabla' f(\nu)\Psi\nabla f(\nu)),
\end{equation}
where $\nabla f$ stands for the derivative of $f$.

The standard error for such a test statistic $\hat{f}$ then amounts to:
\begin{equation}
  SE(\hat{f}) = \sqrt{\frac{\nabla' f(\nu)\Psi\nabla f(\nu)}{T-M}},
\end{equation}
so we require a consistent estimator $\hat{\Psi}$ for $\Psi$.

The standard method to provide such an esimator is to apply
heteroskedasticity- and autocorrelation-robust kernel estimation
to obtain the estimate
\begin{equation}
  \Psi_{T-M} = \frac{T-M}{T-M-4}\sum_{j = -T+M+1}^{T-M-1}Ker\left(
    \frac{j}{S_{T-M}}\right) \hat{\Gamma}_{T-M}(j),
\end{equation}
where a kernel function $Ker(\cdot)$ and a bandwidth $S_{T-M}$ need to be
chosen.


Then a two-sided $p$-value for the hypothesis $H_0$: $f = 0$ is given as:
\begin{equation}
  \hat{p} = 2\Phi\frac{\lvert\hat{f}\rvert}{SE(\hat{f})}.
\end{equation}

%% file: App_Descr_stat.tex
\begin{figure}[h!]
  \includegraphics[scale=0.8]{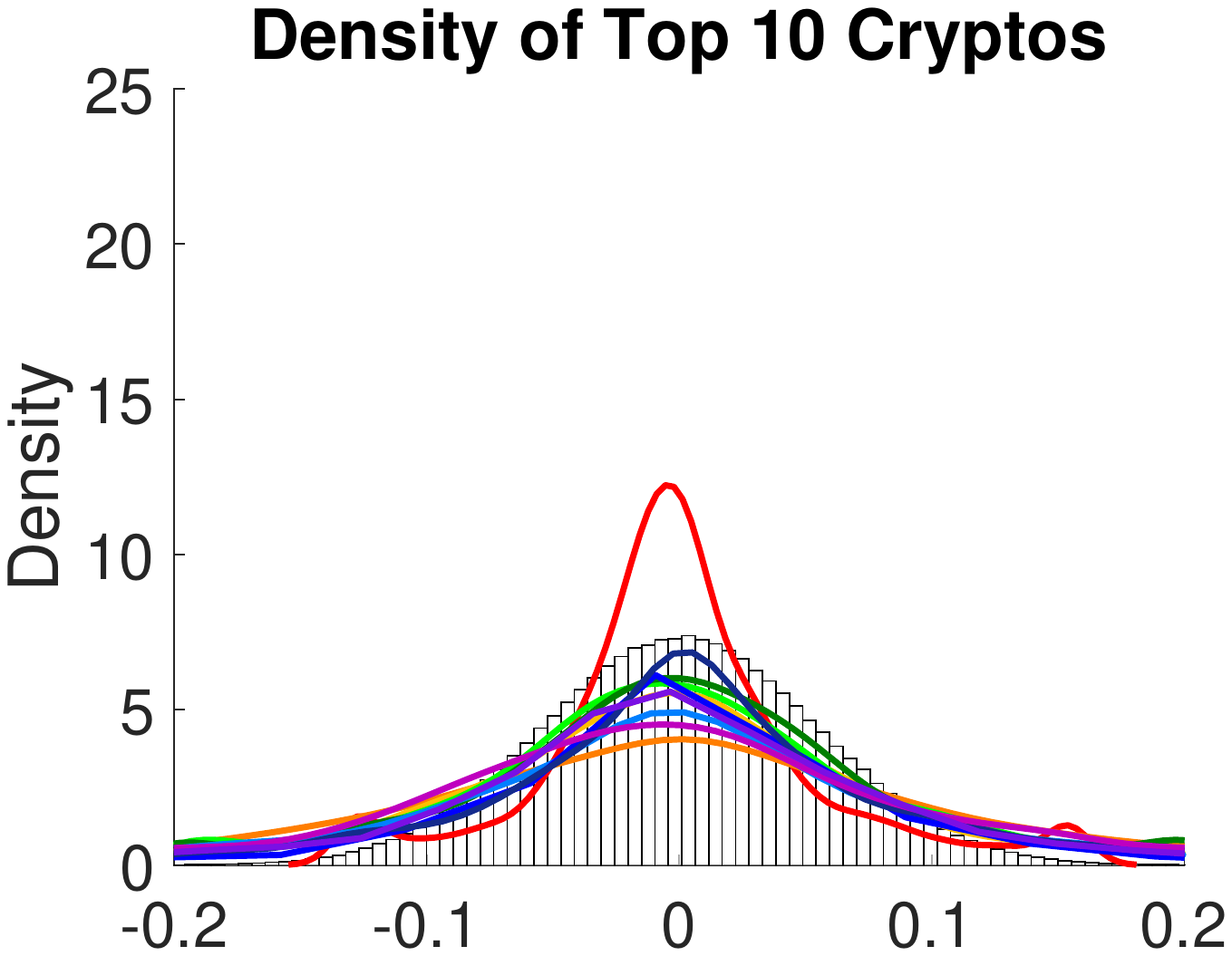}
  \vspace*{-80mm}
  \caption[Density of daily log returns of the top 10 CCs]{Density of daily
    log returns of the top 10 CCs (\textcolor[rgb]{1,0,0}{DOGE},
    \textcolor[rgb]{1,0.494,0}{ZET}, \textcolor[rgb]{1,0.749,0}{XMG},
    \textcolor[rgb]{0,1,0}{SYS}, \textcolor[rgb]{0,0.498,0}{POT},
    \textcolor[rgb]{0,0.498,1}{DGC}, \textcolor[rgb]{0,0,1}{DMD},
    \textcolor[rgb]{0.0784,0.1686,0.5490}{RBY},
    \textcolor[rgb]{0.749,0,0.749}{START},
    \textcolor[rgb]{0.478,0.0627,0.8941}{EMC2}) against a normal distribution
    with same mean and variance. The time span is \periodRange.}
  \label{ReturnsDensity}
  \quantlet{CCPEfficient_surface}{CCPHistReturnsDensity}
\end{figure}

For completeness, we present descriptive statistics both for our traditional
assets as well as all \XsecN\ CCs in our sample.  \reftab{cor-sas} shows
that, as expected, correlations between CCs and traditional assets are low to
non-existent, also in our sample.  \reftabs{DescrstatIND}{DescrstatCC} show
univariate distributional properties of daily log returns on traditional
assets and CCs, respectively.  The generally elevated magnitude for CCs is
clear; \reffig{ReturnsDensity} visually confirms the strong leptokurtic
nature of CC returns.

\tab{cor-sas}{Correlation coefficients of daily log returns of the top ten
  CCs with all conventional financial assets in our analysis (detailed in
  \reftab{list_assets}) over the entire sample period \periodRange.}

\tab{DescrstatIND}{Descriptive statistics for daily log returns (in \%) of
  all conventional assets in our baseline portfolio (detailed in
  \reftab{list_assets}) over the entire sample period \periodRange.
  \descrstatvars}

%

\tab{DescrstatCC}{Descriptive statistics for daily log returns (in \%) of all
  \XsecN\ CCs eligible for our portfolio strategies (detailed in
  \reftab{list_strategies}) over the entire sample period \periodRange.
  \descrstatvars}

%% file: App_RiskContrib.tex

The outcome of portfolio optimization can be viewed in two different ways:
first, in terms of the weights the chosen strategy assigns to each asset;
second, in terms of the risk each constituent contributes to the portfolio.
While flip sides of the same coin, with strongly divergent statistical
properties across assets, as in our case, relative risk contributions can
differ noticably from relative portfolio shares.  For instance, if a
portfolio were to hold the same percentage of its value in UK bonds and in
bitcoin, the changes in portfolio value over time driven by BTC will amount
to a multiple of those stemming from the same-sized fixed-income position.

\begin{figure}[ht!]
  \hspace{-13mm}
  \includegraphics[scale=0.9]{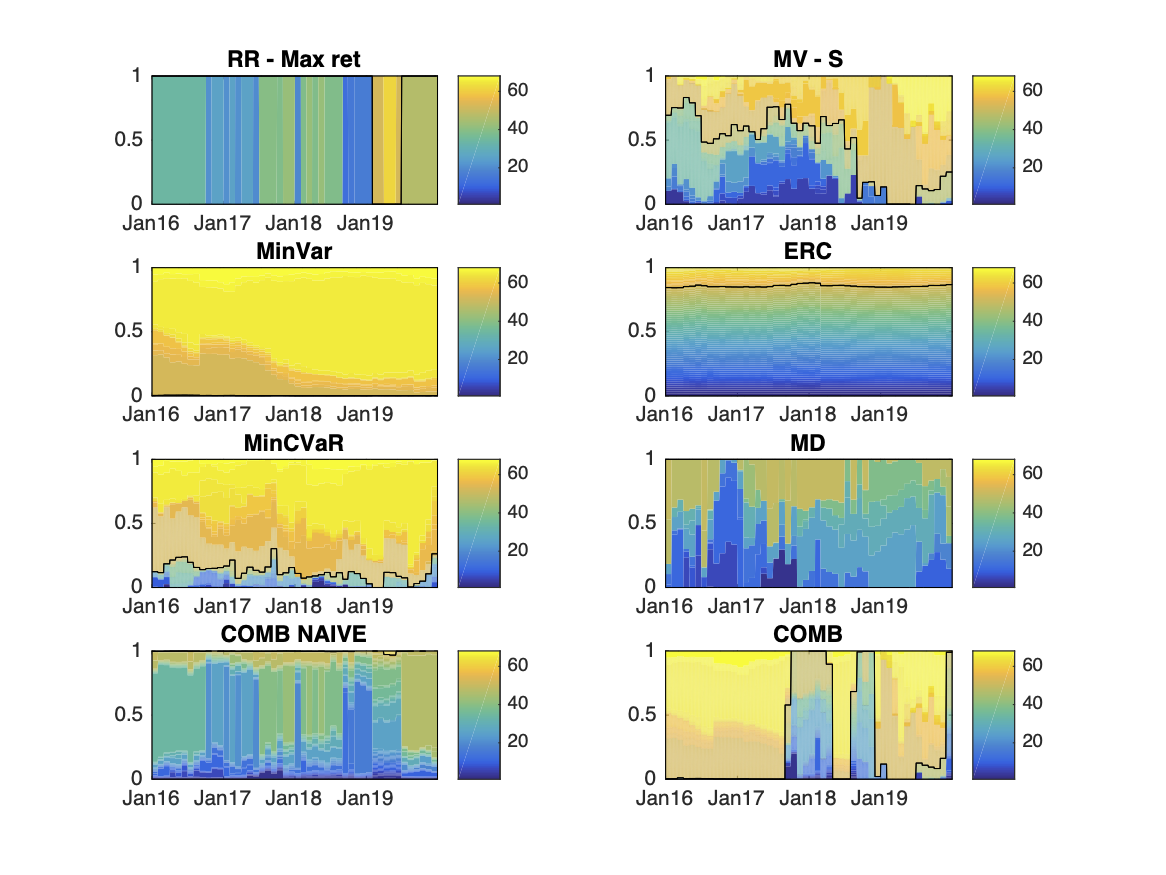}
  \vspace{-7mm}
  \caption[Risk contributions unconstrained]{Evolution of \emph{risk
      contributions} (i.e., fraction of portfolio value changes driven by
    each constituent) of all allocation strategies (without liquidity
    constraints) with monthly rebalancing over the period \periodRange:
    \blacklineseparates.}
  \label{Risk_LIBRO_no}
  \quantlet{CCPRisk_contribution}{CCPRisk\_contribution}
\end{figure}

While we reported weigts in \reffigs{Weights_LIBRO_no}{Weights_LIBRO_yes} in
the main text, for completeness we present the risk contributions as a
function of time in \reffigs{Risk_LIBRO_no}{Risk_LIBRO_yes} for portfolio
optimizations without and with enforced liquidity constraints, respectively.

\begin{figure}[ht!]
  \hspace{-13mm}
  \includegraphics[scale=0.9]{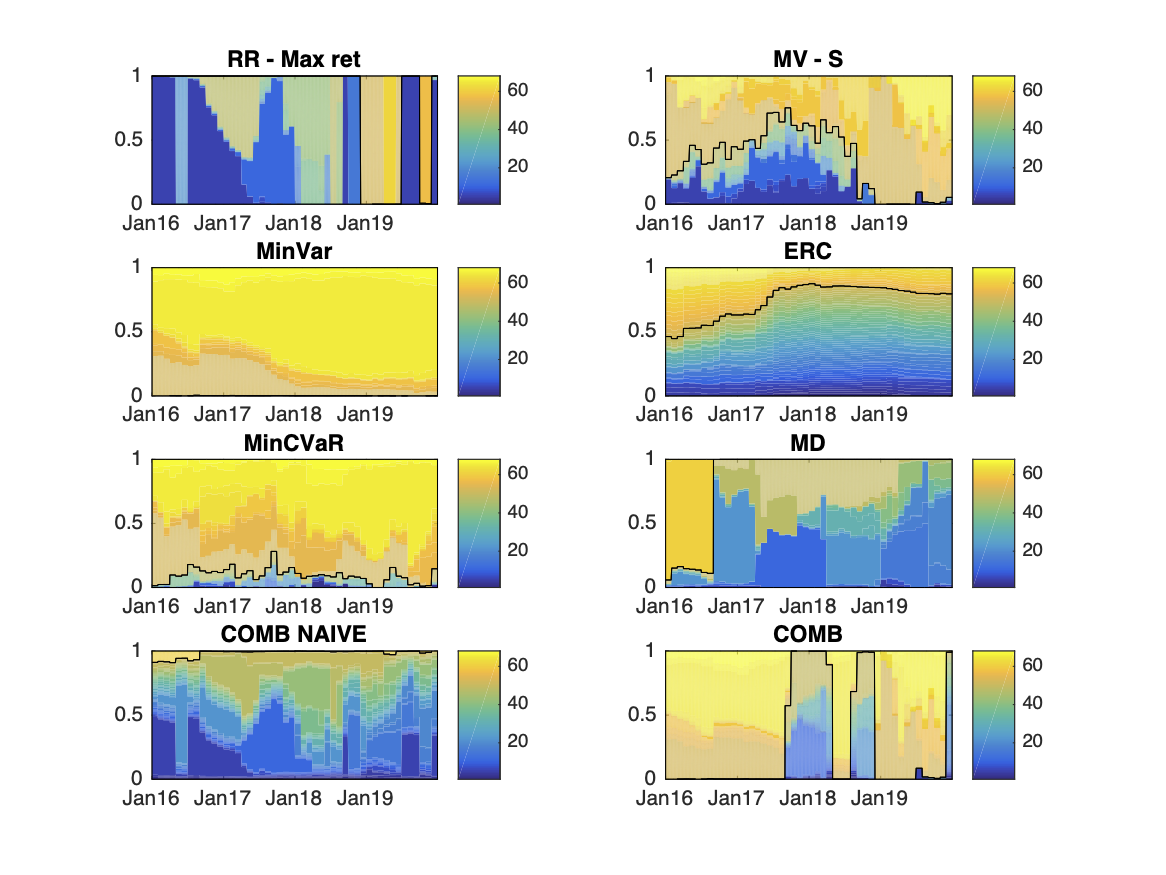}
  \vspace{-7mm}
  \caption[Risk contributions LIBRO]{Evolution of \emph{risk contributions}
    (i.e., fraction of portfolio value changes driven by each constituent) of
    all allocation strategies \withlibrolimits\ with monthly rebalancing over
    the period \periodRange: \blacklineseparates.}
  \label{Risk_LIBRO_yes}
  \quantlet{CCPRisk_contribution}{CCPRisk\_contribution}
\end{figure}

%% file: App_Performance.tex
While our main analysis maintained the industry standard of rebalancing on a
monthly basis, we deem it important to also consider higher trading
frequencies in the CC market.  We therefore report the performance results
based on weekly rebalancing in \reftab{Performance_weekly}, as well as for
daily reallocations in \reftab{Performance_daily}.

Since diversification effects can also be affected by the rebalancing
frequencies, \reftabs{Diversification_weekly}{Diversification_daily} display
the diversification measures for a weekly and daily frequency, respectively.

\tab{Performance_weekly}{\perfmeasures, with \emph{weekly rebalancing} ($k =
  5$). \measuresAre\ \trAmeans\ Strategies are detailed in
  \reftab{list_strategies}.   \superiorperfred\ } %

\tab{Diversification_weekly}{\diversifmeasures\ for \emph{weekly rebalancing}
  ($k = 5$). \diversifmeasuresexplained } \superiorperfred\  

\tab{Performance_daily}{\perfmeasures, with \emph{daily rebalancing} ($k =
  1$).  \measuresAre\ \trAmeans\ Strategies are detailed in
  \reftab{list_strategies}.  \superiorperfred\  }

\tab{Diversification_daily}{\diversifmeasures\ for \emph{daily rebalancing}
  ($k = 1$). \diversifmeasuresexplained  \superiorperfred\  } 

%% file: alla 2/bibliography.bib
@ARTICLE{avramov2002stock,
  Title                    = {Stock return predictability and model uncertainty},
  Author                   = {Avramov, Doron},
  Journal                  = {Journal of Financial Economics},
  Year                     = {2002},
  Number                   = {3},
  Pages                    = {423--458},
  Volume                   = {64},

  Publisher                = {Elsevier}
}

@ARTICLE{briere_virtual_2015,
  author  = {Bri{\`e}re, Marie and Oosterlinck, Kim and Szafarz, Ariane},
  title   = {Virtual currency, tangible return: {Portfolio} diversification with bitcoin},
  journal = {Journal of Asset Management},
  year    = {2015},
  volume  = {16},
  number  = {6},
  pages   = {365--373},
}

@ARTICLE{chopra1993effect,
  title={The Effect of Errors in Means, Variances, and Covariances on Optimal Portfolio Choice},
  author={Chopra, Vijay Kumar and Ziemba, William T},
  journal={The Journal of Portfolio Management},
  volume={19},
  number={2},
  pages={6--11},
  year={1993},
  publisher={Institutional Investor Journals Umbrella}
}

@ARTICLE{choueifaty2008toward,
  Title                    = {Toward maximum diversification},
  Author                   = {Choueifaty, Yves and Coignard, Yves},
  Journal                  = {Journal of Portfolio Management},
  Year                     = {2008},
  Number                   = {1},
  Pages                    = {40--51},
  Volume                   = {35},

  Publisher                = {Euromoney Institutional Investor PLC}
}

@ARTICLE{choueifaty2011properties,
  Title    = {Properties of the most diversified portfolio},
  Author   = {Choueifaty, Yves and Froidure, Tristan and Reynier, Julien},
  journal  = {Journal of Investment Strategies},
  volume   = {2},
  number   = {2},
  pages    = {49--70},
  Year     = {2011}
}

@ARTICLE{chow2011survey,
  title={A survey of alternative equity index strategies},
  author={Chow, {Tzee-man} and Hsu, Jason and Kalesnik, Vitali and Little, Bryce},
  journal={Financial Analysts Journal},
  volume={67},
  number={5},
  pages={37--57},
  year={2011},
  publisher={CFA Institute}
}

@ARTICLE{chuen2017cryptocurrency,
  Title    = {Cryptocurrency: {A} New Investment Opportunity?},
  Author   = {Chuen, David LEE Kuo and Guo, Li and Wang, Yu},
  Journal  = {The Journal of Alternative Investments},
  Year     = {2017},
  Number   = {3},
  Pages    = {16--40},
  Volume   = {20},

  Publisher = {Institutional Investor Journals Umbrella}
}

@ARTICLE{clarke2013risk,
  Title                    = {Risk parity, maximum diversification, and minimum variance: An analytic perspective},
  Author                   = {Clarke, Roger and De Silva, Harindra and Thorley, Steven},
  Journal                  = {Journal of Portfolio Management},
  Year                     = {2013},
  Number                   = {3},
  Pages                    = {39--53},
  Volume                   = {39},

  Publisher                = {Euromoney Institutional Investor PLC}
}

@ARTICLE{clemen1989combining,
  Title                    = {Combining forecasts: {A} review and annotated bibliography},
  Author                   = {Clemen, Robert T},
  Journal                  = {International Journal of Forecasting},
  Year                     = {1989},
  Number                   = {4},
  Pages                    = {559--583},
  Volume                   = {5},

  Publisher                = {Elsevier}
}

@ARTICLE{demiguel2009optimal,
  Title                    = {Optimal versus naive diversification: {How} inefficient is the 1/{N} portfolio strategy?},
  Author                   = {DeMiguel, Victor and Garlappi, Lorenzo and Uppal, Raman},
  Journal                  = {The Review of Financial Studies},
  Year                     = {2009},
  Number                   = {5},
  Pages                    = {1915--1953},
  Volume                   = {22},

  Publisher                = {Oxford University Press}
}

@UNPUBLISHED{eisl_caveat_2015,
  author      = {Eisl, Alexander and Gasser, Stephan M. and Weinmayer, Karl},
  title       = {Caveat Emptor: {Does} {Bitcoin} Improve Portfolio Diversification?},
  institution = {Social Science Research Network},
  year        = {2015},
  type        = {{SSRN} {Scholarly} {Paper}},
  number      = {ID 2408997},
  address     = {Rochester, NY},
  note        = {Available at SSRN 2408997},
}

@InCollection{elendner_cross-section_2017,
  author    = {Elendner, Hermann and Trimborn, Simon and Ong, Bobby and Lee, Teik Ming},
  title     = {The {Cross}-{Section} of {Crypto}-{Currencies} as {Financial} {Assets}},
  booktitle = {Handbook of {Digital} {Finance} and {Financial} {Inclusion}: {Cryptocurrency}, {FinTech}, {InsurTech}, and {Regulation}},
  publisher = {Academic Press},
  year      = {2018},
  editor    = {Lee Kuo Chuen, David and Deng, Robert},
  volume    = {1},
}

@InCollection{hardle_applications_2015,
  author    = {H{\"a}rdle, Wolfgang Karl and Simar, L{\'e}opold},
  title     = {Applications in {Finance}},
  booktitle = {Applied {Multivariate} {Statistical} {Analysis}},
  publisher = {Springer},
  year      = {2015},
  pages     = {487--499},
  edition   = {4},
}

@ARTICLE{Hafner2018,
    author = {Hafner, Christian M},
    title = "{Testing for Bubbles in Cryptocurrencies with Time-Varying Volatility}",
    journal = {Journal of Financial Econometrics},
    volume = {18},
    number = {2},
    pages = {233--249},
    year = {2018},
    month = {10},
    issn = {1479-8409},
    doi = {10.1093/jjfinec/nby023},
    url = {https://doi.org/10.1093/jjfinec/nby023},
    eprint = {https://academic.oup.com/jfec/article-pdf/18/2/233/33218303/nby023.pdf},
}

@ARTICLE{haugen1991efficient,
  title={The efficient market inefficiency of capitalization-weighted stock portfolios},
  author={Haugen, Robert A and Baker, Nardin L},
  journal={The Journal of Portfolio Management},
  volume={17},
  number={3},
  pages={35--40},
  year={1991},
  publisher={Institutional Investor Journals Umbrella}
}

@ARTICLE{huberman1987mean,
  Title                    = {Mean-variance spanning},
  Author                   = {Huberman, Gur and Kandel, Shmuel},
  Journal                  = {The Journal of Finance},
  Year                     = {1987},
  Number                   = {4},
  Pages                    = {873--888},
  Volume                   = {42},

  Publisher                = {Wiley Online Library}
}

@ARTICLE{jagannathan_risk_2003,
  author  = {Jagannathan, Ravi and Ma, Tongshu},
  title   = {Risk Reduction in Large Portfolios: {Why} Imposing the Wrong Constraints Helps},
  journal = {The Journal of Finance},
  year    = {2003},
  volume  = {58},
  number  = {4},
  pages   = {1651--1683},
  issn    = {1540-6261},
}

@ARTICLE{jobson1981performance,
  Title     = {Performance hypothesis testing with the {Sharpe} and {Treynor} measures},
  Author    = {Jobson, J Dave and Korkie, Bob M},
  Journal   = {The Journal of Finance},
  Year      = {1981},
  Number    = {4},
  Pages     = {889--908},
  Volume    = {36},

  Publisher = {Wiley Online Library}
}

@ARTICLE{jorion1985international,
  title={International portfolio diversification with estimation risk},
  author={Jorion, Philippe},
  journal={Journal of Business},
  volume = {58},
  issue = {3},
  pages={259--278},
  year={1985},
  publisher={JSTOR}
}

@ARTICLE{kan2012tests,
  Title     = {Tests of Mean-Variance Spanning},
  Author    = {Kan, Raymond and Zhou, GuoFu},
  Journal   = {Annals of Economics \& Finance},
  Year      = {2012},
  Number    = {1},
  Volume    = {13},
  pages     = {145--193}
}

@article{kan2007optimal,
  title={Optimal portfolio choice with parameter uncertainty},
  author={Kan, Raymond and Zhou, Guofu},
  journal={Journal of Financial and Quantitative Analysis},
  volume={42},
  number={3},
  pages={621--656},
  year={2007},
  publisher={Cambridge University Press}
}

@ARTICLE{krokhmal2002portfolio,
  Title                    = {Portfolio optimization with conditional value-at-risk objective and constraints},
  Author                   = {Krokhmal, Pavlo and Palmquist, Jonas and Uryasev, Stanislav},
  Journal                  = {Journal of Risk},
  Year                     = {2002},
  Pages                    = {43--68},
  Volume                   = {4}
}

@ARTICLE{ledoit2008robust,
  Title                    = {Robust performance hypothesis testing with the {Sharpe} ratio},
  Author                   = {Ledoit, Oliver and Wolf, Michael},
  Journal                  = {Journal of Empirical Finance},
  Year                     = {2008},
  Number                   = {5},
  Pages                    = {850--859},
  Volume                   = {15},

  Publisher                = {Elsevier}
}

@ARTICLE{maillard2010properties,
  author  = {Maillard, S{\'e}bastien and Roncalli, Thierry and Te{\"i}letche, J{\'e}r{\^o}me},
  title   = {The properties of equally weighted risk contribution portfolios},
  journal = {The Journal of Portfolio Management},
  year    = {2010},
  volume  = {36},
  number  = {4},
  pages   = {60--70},
}

@ARTICLE{markowitz_portfolio_1952,
  author  = {Markowitz, Harry},
  title   = {Portfolio Selection},
  journal = {The Journal of Finance},
  year    = {1952},
  volume  = {7},
  number  = {1},
  pages   = {77--91},
  issn    = {0022-1082},
  doi     = {10.2307/2975974},
  urldate = {2017-04-27},
}

@ARTICLE{merton1980estimating,
  Title                    = {On estimating the expected return on the market: {An} exploratory investigation},
  Author                   = {Merton, Robert C},
  Journal                  = {Journal of Financial Economics},
  Year                     = {1980},
  Number                   = {4},
  Pages                    = {323--361},
  Volume                   = {8},

  Publisher                = {Elsevier}
}

@ARTICLE{pezier2006relative,
  author  = {Pezier, Jacques and White, Anthony},
  title   = {The relative merits of investable hedge fund indices and of funds of hedge funds in optimal passive portfolios},
  journal = {The Journal of Alternative Investments},
  year    = {2008},
  volume  = {10},
  number  = {4},
  pages   = {37-49},
  doi     = {https://doi.org/10.3905/jai.2008.705531},
}

@ARTICLE{politis1994stationary,
  Title                    = {The stationary bootstrap},
  Author                   = {Politis, Dimitris N. and Romano, Joseph P.},
  Journal                  = {Journal of the American Statistical Association},
  Year                     = {1994},
  Number                   = {428},
  Pages                    = {1303--1313},
  Volume                   = {89},

  Publisher                = {Taylor \& Francis Group}
}

@ARTICLE{politis2004automatic,
  Title                    = {Automatic block-length selection for the dependent bootstrap},
  Author                   = {Politis, Dimitris N. and White, Halbert},
  Journal                  = {Econometric Reviews},
  Year                     = {2004},
  Number                   = {1},
  Pages                    = {53--70},
  Volume                   = {23},

  Publisher                = {Taylor \& Francis}
}

@ARTICLE{qian2005financial,
  Title                    = {On the financial interpretation of risk contribution: {R}isk budgets do add up},
  Author                   = {Qian, Edward E},
  Year                     = {2006},
  Journal                  = {Journal of Investment Management},
  Volume                   = {4},
  Number                   = {4},
  Pages                    = {41--51},
}

@ARTICLE{rockafellar2000optimization,
  Title                    = {Optimization of conditional value-at-risk},
  Author                   = {Rockafellar, R. Tyrrell and Uryasev, Stanislav},
  Journal                  = {Journal of Risk},
  Year                     = {2000},
  Pages                    = {21--42},
  Volume                   = {2}
}

@ARTICLE{rudin2006portfolio,
  Title                    = {A portfolio diversification index},
  Author                   = {Rudin, Alexander M. and Morgan, Jonathan S},
  Journal                  = {The Journal of Portfolio Management},
  Year                     = {2006},
  Number                   = {2},
  Pages                    = {81--89},
  Volume                   = {32},

  Publisher                = {Institutional Investor Journals}
}

@Article{Scaillet2018,
    author = {Scaillet, Olivier and Treccani, Adrien and Trevisan, Christopher},
    title = "{High-Frequency Jump Analysis of the Bitcoin Market}",
    journal = {Journal of Financial Econometrics},
    volume = {18},
    number = {2},
    pages = {209--232},
    year = {2018},
    month = {06},
    issn = {1479-8409},
    doi = {10.1093/jjfinec/nby013},
    url = {https://doi.org/10.1093/jjfinec/nby013},
    eprint = {https://academic.oup.com/jfec/article-pdf/18/2/209/33218377/nby013.pdf},
}

@ARTICLE{schanbacher2015averaging,
  Title                    = {Averaging across asset allocation models},
  Author                   = {Schanbacher, Peter},
  Journal                  = {Jahrb{\"u}cher f{\"u}r National{\"o}konomie und Statistik},
  Year                     = {2015},
  Number                   = {1},
  Pages                    = {61--81},
  Volume                   = {235},

  Publisher                = {Lucius \& Lucius}
}

@ARTICLE{schanbacher2014combining,
  author    = {Schanbacher, Peter},
  title     = {Combining Portfolio Models},
  journal   = {Annals of Economics and Finance},
  year      = {2014},
  volume    = {15},
  number    = {2},
  pages     = {433--455},
  publisher = {WUHAN UNIV JOURNALS PRESS DONGHU RD, 155, WUHAN, HUBEI 430072, PEOPLES R CHINA},
}

@ARTICLE{simaan1997estimation,
  title={Estimation risk in portfolio selection: {T}he mean variance model versus the mean absolute deviation model},
  author={Simaan, Yusif},
  journal={Management Science},
  volume={43},
  number={10},
  pages={1437--1446},
  year={1997},
  publisher={INFORMS}
}

@ARTICLE{strongin2000beating,
  Title                    = {Beating benchmarks},
  Author                   = {Strongin, Steve and Petsch, Melanie and Sharenow, Greg},
  Journal                  = {The Journal of Portfolio Management},
  Year                     = {2000},
  Number                   = {4},
  Pages                    = {11--27},
  Volume                   = {26},

  Publisher                = {Institutional Investor Journals Umbrella}
}

@ARTICLE{trimborn_crix_2016,
  author  = {Trimborn, Simon and H{\"a}rdle, Wolfgang Karl},
  title   = {{CRIX} an {Index} for cryptocurrencies},
  journal = {Journal of Empirical Finance},
  Pages   = {107--122},
  Volume  = {49},
  year    = {2018},
}

@ARTICLE{trimborn_investing_2017,
    author  = {Trimborn, Simon and Li, Mingyang and H{\"a}rdle, Wolfgang Karl},
    title = "{Investing with Cryptocurrencies--a Liquidity Constrained Investment Approach}",
    journal = {Journal of Financial Econometrics},
    volume = {18},
    number = {2},
    pages = {280--306},
    year = {2019},
    month = {06},
    issn = {1479-8409},
    doi = {10.1093/jjfinec/nbz016},
    url = {https://doi.org/10.1093/jjfinec/nbz016},
    eprint = {https://academic.oup.com/jfec/article-pdf/18/2/280/33218337/nbz016.pdf},
}

@TechReport{rivin2018cci30,
  author        = {Igor Rivin and Carlo Scevola},
  title         = {The CCI30 Index},
  year          = {2018},
  type          = {White paper},
  note          = {Available at \url{https://arxiv.org/abs/1804.06711}},
  archiveprefix = {arXiv},
  eprint        = {1804.06711},
  primaryclass  = {q-fin.GN},
}

@ARTICLE{walther2018bitcoin,
  author  = {Klein, Tony and Thu, Hien Pham and Walther, Thomas},
  title={Bitcoin is not the New Gold -- {A} comparison of volatility, correlation, and portfolio performance},
  journal={International Review of Financial Analysis},
  volume={59},
  pages={105--116},
  year={2018},
  publisher={Elsevier}
}

@ARTICLE{Shahzad_et_al_2019,
title = "Is Bitcoin a better safe-haven investment than gold and commodities?",
journal = "International Review of Financial Analysis",
volume = "63",
pages = "322--330",
year = "2019",
issn = "1057-5219",
doi = "https://doi.org/10.1016/j.irfa.2019.01.002",
url = "http://www.sciencedirect.com/science/article/pii/S1057521918306604",
author = "Syed Jawad Hussain Shahzad and Elie Bouri and David Roubaud and Ladislav Kristoufek and Brian Lucey"
}

@ARTICLE{Dyhrberg_2016,
title = "Hedging capabilities of bitcoin. Is it the virtual gold?",
journal = "Finance Research Letters",
volume = "16",
pages = "139--144",
year = "2016",
issn = "1544-6123",
doi = "https://doi.org/10.1016/j.frl.2015.10.025",
url = "http://www.sciencedirect.com/science/article/pii/S1544612315001208",
author = "Anne Haubo Dyhrberg"
}

@PHDTHESIS{wyss_measuring_2004,
  author  = {von Wyss, Rico},
  title   = {Measuring and {Predicting} {Liquidity} in the {Stock} {Market}},
  school  = {Universit\"at St.\ Gallen}, 
  year    = {2004},
}

@ARTICLE{Alessandretti2018,
  doi = {10.1155/2018/8983590},
  url = {https://doi.org/10.1155/2018/8983590},
  year = {2018},
  month = nov,
  publisher = {Hindawi Limited},
  volume = {2018},
  pages = {1--16},
  author = {Laura Alessandretti and Abeer {ElBahrawy} and Luca Maria Aiello and Andrea Baronchelli},
  title = {Anticipating Cryptocurrency Prices Using Machine Learning},
  journal = {Complexity}
}

@TechReport{Nakamoto_2008,
  Title                    = {Bitcoin: A peer-to-peer electronic cash system},
  Author                   = {Nakamoto, Satoshi},
  Type                     = {White paper},
  note                     = {Available at: \url{https://bitcoin.org/bitcoin.pdf}},
  Year                     = {2008},
  Url                      = {https://bitcoin.org/bitcoin.pdf}
}

@TechReport{crypto20,
  author = {Daniel Schwartzkopff and Luke Schwartzkopff and Raymond Botha and Matthew Finlayson and Frans Cronje},
  title  = {CRYPTO20: The First Tokenized Cryptocurrency Index Fund},
  year   = {2017},
  type   = {White paper},
  note   = {Available at: \url{https://cdn.crypto20.com/pdf/c20-whitepaper.pdf}},
  url    = {https://cdn.crypto20.com/pdf/c20-whitepaper.pdf},
}

@InCollection{Yermack_2015,
  author    = {David Yermack},
  title     = {{I}s Bitcoin a Real Currency? {A}n Economic Appraisal},
  booktitle = {Handbook of Digital Currency},
  publisher = {Academic Press},
  year      = {2015},
  editor    = {Chuen, David Lee Kuo},
  chapter   = {2},
  pages     = {31--43},
  address   = {San Diego},
  isbn      = {978-0-12-802117-0},
  doi       = {http://dx.doi.org/10.1016/B978-0-12-802117-0.00002-3},
  url       = {http://www.sciencedirect.com/science/article/pii/B9780128021170000023},
}

@ARTICLE{Carhart_1997,
 ISSN = {00221082, 15406261},
 URL = {http://www.jstor.org/stable/2329556},
 author = {Mark M.\ Carhart},
 journal = {The Journal of Finance},
 number = {1},
 pages = {57--82},
 publisher = {[American Finance Association, Wiley]},
 title = {On Persistence in Mutual Fund Performance},
 volume = {52},
 year = {1997}
}

@ARTICLE{Jegadeesh_Titman_1993,
 ISSN = {00221082, 15406261},
 URL = {http://www.jstor.org/stable/2328882},
 author = {Narasimhan Jegadeesh and Sheridan Titman},
 journal = {The Journal of Finance},
 number = {1},
 pages = {65--91},
 publisher = {[American Finance Association, Wiley]},
 title = {Returns to Buying Winners and Selling Losers: Implications for Stock Market Efficiency},
 volume = {48},
 year = {1993}
}

@ARTICLE{Fama_French_1992,
author = {Eugene Fama and Kenneth French},
journal = {The Journal of Finance},
number = {2},
pages = {427--465},
publisher = {[American Finance Association, Wiley]},
title = {The Cross-Section of Expected Stock Returns},
volume = {47},
year = {1992},
doi = {10.1111/j.1540-6261.1992.tb04398.x},
url = {https://onlinelibrary.wiley.com/doi/abs/10.1111/j.1540-6261.1992.tb04398.x},
eprint = {https://onlinelibrary.wiley.com/doi/pdf/10.1111/j.1540-6261.1992.tb04398.x}
}

@article{Fama_French_1993,
title = "Common risk factors in the returns on stocks and bonds",
journal = "Journal of Financial Economics",
volume = "33",
number = "1",
pages = "3--56",
year = "1993",
issn = "0304-405X",
doi = "https://doi.org/10.1016/0304-405X(93)90023-5",
url = "http://www.sciencedirect.com/science/article/pii/0304405X93900235",
author = "Eugene F. Fama and Kenneth R. French"
}

@ARTICLE{Fama_French_2015,
title = "A five-factor asset pricing model",
journal = "Journal of Financial Economics",
volume = "116",
number = "1",
pages = "1--22",
year = "2015",
issn = "0304-405X",
doi = "https://doi.org/10.1016/j.jfineco.2014.10.010",
url = "http://www.sciencedirect.com/science/article/pii/S0304405X14002323",
author = "Eugene F. Fama and Kenneth R. French"
}

@ARTICLE{Ross_1976,
title = "The arbitrage theory of capital asset pricing",
journal = "Journal of Economic Theory",
volume = "13",
number = "3",
pages = "341--360",
year = "1976",
issn = "0022-0531",
doi = "https://doi.org/10.1016/0022-0531(76)90046-6",
url = "http://www.sciencedirect.com/science/article/pii/0022053176900466",
author = "Stephen A Ross"
}

@UNPUBLISHED{Liu_Tsyvinski_2018,
  author      = {Liu, Yukun and Tsyvinski, Aleh},
  title       = {Risks and Returns of Cryptocurrency},
  institution = {Social Science Research Network},
  year        = {2018},
  type        = {{SSRN} {Scholarly} {Paper}},
  number      = {ID 3226952},
  address     = {Rochester, NY},
  note   = {Available at SSRN 3226952},
}

@UNPUBLISHED{Liu_Tsyvinski_2019,
  author      = {Liu, Yukun and Tsyvinski, Aleh and Wu, Xi},
  title       = {Common Risk Factors in Cryptocurrency},
  institution = {Social Science Research Network},
  year        = {2019},
  type        = {{SSRN} {Scholarly} {Paper}},
  number      = {ID 3379131},
  address     = {Rochester, NY},
  note   = {Available at SSRN 3379131},
}

@ARTICLE{Platanakis_Urquhart_2019,
title = "Should investors include Bitcoin in their portfolios? {A} portfolio theory approach",
journal = "The British Accounting Review",
pages = "100837",
year = "2019",
issn = "0890-8389",
doi = "https://doi.org/10.1016/j.bar.2019.100837",
url = "http://www.sciencedirect.com/science/article/pii/S0890838919300605",
author = "Emmanouil Platanakis and Andrew Urquhart"
}

@ARTICLE{Akhtaruzzaman_Sensoy_Corbet_2019,
title = "The influence of Bitcoin on portfolio diversification and design",
journal = "Finance Research Letters",
pages = "101344",
year = "2019",
issn = "1544-6123",
doi = "https://doi.org/10.1016/j.frl.2019.101344",
url = "http://www.sciencedirect.com/science/article/pii/S1544612319305215",
author = "Md Akhtaruzzaman and Ahmet Sensoy and Shaen Corbet"
}

@ARTICLE{Boehme_et_al_2015-JEP,
Author = {B\"ohme, Rainer and Christin, Nicolas and Edelman, Benjamin and Moore, Tyler},
Title = {Bitcoin: Economics, Technology, and Governance},
Journal = {Journal of Economic Perspectives},
Volume = {29},
Number = {2},
Year = {2015},
Month = {5},
Pages = {213--38},
DOI = {10.1257/jep.29.2.213},
URL = {http://www.aeaweb.org/articles?id=10.1257/jep.29.2.213}
}

@UNPUBLISHED{Hubrich_2017,
  author      = {Hubrich, Stefan},
  title       = {Know when to hodl them, know when to fodl them: {A}n Investigation of Factor Based Investing in the Cryptocurrency Space},
  year        = {2017},
  number      = {ID 3055498},
  month       = {10},
  address     = {Rochester, NY},
  doi         = {10.13140/RG.2.2.35090.96969},
  type        = {{SSRN} {Scholarly} {Paper}},
  note   = {Available at SSRN 3055498},
}

@ARTICLE{Wang_Vergne_2017,
    author = {Wang, Sha and Vergne, Jean-Philippe},
    journal = {PLOS ONE},
    publisher = {Public Library of Science},
    title = {Buzz Factor or Innovation Potential: What Explains Cryptocurrencies’ Returns?},
    year = {2017},
    month = {01},
    volume = {12},
    url = {https://doi.org/10.1371/journal.pone.0169556},
    pages = {1--17},
    number = {1},
    doi = {10.1371/journal.pone.0169556}
}

@ARTICLE{Brauneis_Mestel_2018,
title = "Price discovery of cryptocurrencies: Bitcoin and beyond",
journal = "Economics Letters",
volume = "165",
pages = "58--61",
year = "2018",
issn = "0165-1765",
doi = "https://doi.org/10.1016/j.econlet.2018.02.001",
url = "http://www.sciencedirect.com/science/article/pii/S0165176518300417",
author = "Alexander Brauneis and Roland Mestel"
}

@ARTICLE{Zhang_Wang_Li_Shen_2018,
author = {Wei Zhang and Pengfei Wang and Xiao Li and Dehua Shen},
title = {Some stylized facts of the cryptocurrency market},
journal = {Applied Economics},
volume = {50},
number = {55},
pages = {5950--5965},
year  = {2018},
publisher = {Routledge},
doi = {10.1080/00036846.2018.1488076},
URL = {https://doi.org/10.1080/00036846.2018.1488076},
eprint = {https://doi.org/10.1080/00036846.2018.1488076}
}

@TechReport{Elendner_2018,
author = {Hermann Elendner},
year = {2018},
month = {09},
title = {F5: optimised crypto-currency investment strategies},
type = {White paper},
note = {Available at \url{f5crypto.com}},
url = {\url{f5crypto.com}}
}

@ARTICLE{Sovbetov_2018,
author = {Sovbetov, Yhlas},
year = {2018},
month = {02},
pages = {1--27},
title = {Factors Influencing Cryptocurrency Prices: Evidence from Bitcoin, Ethereum, Dash, Litcoin, and Monero},
volume = {2},
journal = {Journal of Economics and Financial Analysis}
}

@ARTICLE{Urquhart_2017,
title = "Price clustering in Bitcoin",
journal = "Economics Letters",
volume = "159",
pages = "145--148",
year = "2017",
issn = "0165-1765",
doi = "https://doi.org/10.1016/j.econlet.2017.07.035",
url = "http://www.sciencedirect.com/science/article/pii/S0165176517303233",
author = "Andrew Urquhart"
}

@ARTICLE{Blau_2018,
title = "Price dynamics and speculative trading in Bitcoin",
journal = "Research in International Business and Finance",
volume = "43",
pages = "15--21",
year = "2018",
issn = "0275-5319",
doi = "https://doi.org/10.1016/j.ribaf.2017.07.183",
url = "http://www.sciencedirect.com/science/article/pii/S0275531917304750",
author = "Benjamin M. Blau"
}

@ARTICLE{Chu_Nadarajah_Chan_2015,
    author = {Chu, Jeffrey and Nadarajah, Saralees and Chan, Stephen},
    journal = {PLOS ONE},
    publisher = {Public Library of Science},
    title = {Statistical Analysis of the Exchange Rate of Bitcoin},
    year = {2015},
    month = {07},
    volume = {10},
    url = {https://doi.org/10.1371/journal.pone.0133678},
    pages = {1-27},
    number = {7},
    doi = {10.1371/journal.pone.0133678}
}

@ARTICLE{Kristoufek_2015,
    author = {Kristoufek, Ladislav},
    journal = {PLOS ONE},
    publisher = {Public Library of Science},
    title = {What Are the Main Drivers of the Bitcoin Price? Evidence from Wavelet Coherence Analysis},
    year = {2015},
    month = {04},
    volume = {10},
    url = {https://doi.org/10.1371/journal.pone.0123923},
    pages = {1-15},
    number = {4},
    doi = {10.1371/journal.pone.0123923}
}

@INPROCEEDINGS{Osterrieder_2017,
  title={The Statistics of Bitcoin and Cryptocurrencies},
  author={Joerg Osterrieder},
  year={2017},
  booktitle={2017 International Conference on Economics, Finance and Statistics (ICEFS 2017)},
  issn={2352-5428},
  isbn={978-94-6252-311-1},
  url={https://doi.org/10.2991/icefs-17.2017.33},
  doi={https://doi.org/10.2991/icefs-17.2017.33},
  publisher={Atlantis Press}
}

@ARTICLE{Bariviera_et_al_2017,
title = "Some stylized facts of the Bitcoin market",
journal = "Physica A: Statistical Mechanics and its Applications",
volume = "484",
pages = "82--90",
year = "2017",
issn = "0378-4371",
doi = "https://doi.org/10.1016/j.physa.2017.04.159",
url = "http://www.sciencedirect.com/science/article/pii/S0378437117304697",
author = "Aurelio F. Bariviera and María José Basgall and Waldo Hasperué and Marcelo Naiouf"
}

@ARTICLE{Cheah_Fry_2015,
title = "Speculative bubbles in Bitcoin markets? An empirical investigation into the fundamental value of Bitcoin",
journal = "Economics Letters",
volume = "130",
pages = "32 - 36",
year = "2015",
issn = "0165-1765",
doi = "https://doi.org/10.1016/j.econlet.2015.02.029",
url = "http://www.sciencedirect.com/science/article/pii/S0165176515000890",
author = "Eng-Tuck Cheah and John Fry"
}

@ARTICLE{Fry_Cheah_2016,
title = "Negative bubbles and shocks in cryptocurrency markets",
journal = "International Review of Financial Analysis",
volume = "47",
pages = "343--352",
year = "2016",
issn = "1057-5219",
doi = "https://doi.org/10.1016/j.irfa.2016.02.008",
url = "http://www.sciencedirect.com/science/article/pii/S1057521916300163",
author = "John Fry and Eng-Tuck Cheah"
}

@ARTICLE{Nunez_et_al_2019,
    author = {Núñez, José Antonio AND Contreras-Valdez, Mario I. AND Franco-Ruiz, Carlos A.},
    journal = {PLOS ONE},
    publisher = {Public Library of Science},
    title = {Statistical analysis of bitcoin during explosive behavior periods},
    year = {2019},
    month = {03},
    volume = {14},
    url = {https://doi.org/10.1371/journal.pone.0213919},
    pages = {1--22},
    number = {3},
    doi = {10.1371/journal.pone.0213919}
}

@ARTICLE{Shen_Urquhart_Wang_2019,
author = {Shen, Dehua and Urquhart, Andrew and Wang, Pengfei},
year = {2019},
month = {07},
issue = {forthcoming},
pages = {},
title = {A three-factor pricing model for cryptocurrencies},
journal = {Finance Research Letters},
doi = {10.1016/j.frl.2019.07.021}
}

@Article{Caginalp_Caginalp_2019,
  author   = {Carey Caginalp and Gunduz Caginalp},
  title    = {Establishing cryptocurrency equilibria through game theory},
  journal  = {AIMS Mathematics},
  year     = {2019},
  volume   = {4},
  number   = {3},
  pages    = {420-436},
  issn     = {2473-6988},
  doi      = {http://dx.doi.org/10.3934/math.2019.3.420},
}

@Article{Bolt_vanOordt_forthc,
  author   = {Bolt, Wilko and {van Oordt}, Maarten R.C.},
  title    = {On the Value of Virtual Currencies},
  journal  = {Journal of Money, Credit and Banking},
  year     = {2020},
  issue   = {forthcoming},
  doi      = {10.1111/jmcb.12619},
  eprint   = {https://onlinelibrary.wiley.com/doi/pdf/10.1111/jmcb.12619},
  url      = {https://onlinelibrary.wiley.com/doi/abs/10.1111/jmcb.12619},
}

@ARTICLE{Houy_2016,
title={The Bitcoin Mining Game},
volume={1},
url={https://ledgerjournal.org/ojs/ledger/article/view/13},
DOI={10.5195/ledger.2016.13},
journal={Ledger},
author={Houy, Nicolas},
year={2016},
month={12},
pages={53--68}
}

@ARTICLE{Dimitri_2017,
title={Bitcoin Mining as a Contest},
volume={2},
url={https://ledgerjournal.org/ojs/ledger/article/view/96},
DOI={10.5195/ledger.2017.96}, 
journal={Ledger},
author={Dimitri, Nicola},
year={2017},
month={9},
pages={31-37}
}

@TECHREPORT{Abadi_Brunnermeier_2019,
       	author = {Joseph Abadi and Markus K. Brunnermeier},
	title = {Blockchain Economics},
	year = {2019},
	type = {Working Paper},
	note = {Available at \url{https://scholar.princeton.edu/markus/publications/blockchain-economics}},
}

@InProceedings{Glaser_et_al_2014,
  author    = {Glaser, Florian and Zimmermann, Kai and Haferkorn, Martin and Weber, Moritz Christian and Siering, Michael},
  title     = {Bitcoin -- Asset or Currency? Revealing Users' Hidden Intentions},
  booktitle = {ECIS 2014},
  year      = {2014},
  type      = {{SSRN} {Scholarly} {Paper}},
}

@ARTICLE{Baur_Hong_Lee_2017,
title = "Bitcoin: Medium of exchange or speculative assets?",
journal = "Journal of International Financial Markets, Institutions and Money",
volume = "54",
pages = "177--189",
year = "2018",
issn = "1042-4431",
doi = "https://doi.org/10.1016/j.intfin.2017.12.004",
url = "http://www.sciencedirect.com/science/article/pii/S1042443117300720",
author = "Dirk G. Baur and KiHoon Hong and Adrian D. Lee"
}

@ARTICLE{Gandal_et_al_2018,
title = "Price manipulation in the Bitcoin ecosystem",
journal = "Journal of Monetary Economics",
volume = "95",
pages = "86--96",
year = "2018",
issn = "0304-3932",
doi = "https://doi.org/10.1016/j.jmoneco.2017.12.004",
url = "http://www.sciencedirect.com/science/article/pii/S0304393217301666",
author = "Neil Gandal and JT Hamrick and Tyler Moore and Tali Oberman"
}

@ARTICLE{Schilling_Uhlig_2019,
title = "Some simple bitcoin economics",
journal = "Journal of Monetary Economics",
volume = "106",
pages = "16--26",
year = "2019",
note = "SPECIAL CONFERENCE ISSUE: “Money Creation and Currency Competition” October 19-20, 2018 Sponsored by the Study Center Gerzensee and Swiss National Bank",
issn = "0304-3932",
doi = "https://doi.org/10.1016/j.jmoneco.2019.07.002",
url = "http://www.sciencedirect.com/science/article/pii/S0304393219301199",
author = "Linda Schilling and Harald Uhlig"
}

@Article{Chen2019,
  author    = {Chen, Cathy Yi-Hsuan and Hafner, Christian M},
  title     = {Sentiment-induced bubbles in the cryptocurrency market},
  journal   = {Journal of Risk and Financial Management},
  year      = {2019},
  volume    = {12},
  number    = {2},
  pages     = {53},
  publisher = {Multidisciplinary Digital Publishing Institute},
}

@Article{Momtaz2019,
  author    = {Momtaz, Paul P},
  title     = {Token sales and initial coin offerings: Introduction},
  journal   = {The Journal of Alternative Investments},
  year      = {2019},
  volume    = {21},
  number    = {4},
  pages     = {7--12},
  publisher = {Institutional Investor Journals Umbrella},
}

@UNPUBLISHED{Kim2019,
  author = {Kim, Alisa and Trimborn, Simon and H{\"a}rdle, Wolfgang K},
  title  = {VCRIX---A Volatility Index for Crypto-Currencies},
  year   = {2019},
  type   = {{SSRN} {Scholarly} {Paper}},
  number = {ID 3480348},
  address= {Rochester, NY},
  note   = {Available at SSRN 3480348},
}

@ARTICLE{Hou_et_al_2020,
    author = {Hou, Ai Jun and Wang, Weining and Chen, Cathy Y H and H\"ardle, Wolfgang Karl},
    title = {Pricing Cryptocurrency Options},
    journal = {Journal of Financial Econometrics},
    volume = {18},
    number = {2},
    pages = {250-279},
    year = {2020},
    month = {05},
    issn = {1479-8409},
    doi = {10.1093/jjfinec/nbaa006},
    url = {https://doi.org/10.1093/jjfinec/nbaa006},
    eprint = {https://academic.oup.com/jfec/article-pdf/18/2/250/33218360/nbaa006.pdf},
}

@UNPUBLISHED{Momtaz2018,
  author = {Momtaz, Paul P},
  title  = {Initial coin offerings},
  year   = {2018},
  type   = {{SSRN} {Scholarly} {Paper}},
  number = {ID 3166709},
  address= {Rochester, NY},
  note   = {Available at SSRN 3166709},
}

@ARTICLE{Wei_2018,
title = "Liquidity and market efficiency in cryptocurrencies",
journal = "Economics Letters",
volume = "168",
pages = "21--24",
year = "2018",
issn = "0165-1765",
doi = "https://doi.org/10.1016/j.econlet.2018.04.003",
url = "http://www.sciencedirect.com/science/article/pii/S0165176518301320",
author = "Wang Chun Wei"
}

@ARTICLE{Lintner_1965,
 ISSN = {00346535, 15309142},
 URL = {http://www.jstor.org/stable/1924119},
 author = {John Lintner},
 journal = {The Review of Economics and Statistics},
 number = {1},
 pages = {13--37},
 publisher = {The MIT Press},
 title = {The Valuation of Risk Assets and the Selection of Risky Investments in Stock Portfolios and Capital Budgets},
 volume = {47},
 year = {1965}
}

@ARTICLE{Sharpe_1964,
 ISSN = {00221082, 15406261},
 URL = {http://www.jstor.org/stable/2977928},
 author = {William F. Sharpe},
 journal = {The Journal of Finance},
 number = {3},
 pages = {425--442},
 publisher = {[American Finance Association, Wiley]},
 title = {Capital Asset Prices: A Theory of Market Equilibrium under Conditions of Risk},
 volume = {19},
 year = {1964}
}

@ARTICLE{Roll_1977,
title = "A critique of the asset pricing theory's tests Part I: On past and potential testability of the theory",
journal = "Journal of Financial Economics",
volume = "4",
number = "2",
pages = "129--176",
year = "1977",
issn = "0304-405X",
doi = "https://doi.org/10.1016/0304-405X(77)90009-5",
url = "http://www.sciencedirect.com/science/article/pii/0304405X77900095",
author = "Richard Roll"
}

@ARTICLE{Cochrane_2011,
author = {Cochrane, John H.},
title = {Presidential Address: Discount Rates},
journal = {The Journal of Finance},
volume = {66},
number = {4},
pages = {1047--1108},
doi = {10.1111/j.1540-6261.2011.01671.x},
url = {https://onlinelibrary.wiley.com/doi/abs/10.1111/j.1540-6261.2011.01671.x},
eprint = {https://onlinelibrary.wiley.com/doi/pdf/10.1111/j.1540-6261.2011.01671.x},
year = {2011}
}

@ARTICLE{NovyMarx_2014,
title = "Predicting anomaly performance with politics, the weather, global warming, sunspots, and the stars",
journal = "Journal of Financial Economics",
volume = "112",
number = "2",
pages = "137--146",
year = "2014",
issn = "0304-405X",
doi = "https://doi.org/10.1016/j.jfineco.2014.02.002",
url = "http://www.sciencedirect.com/science/article/pii/S0304405X14000208",
author = "Robert Novy-Marx"
}

@Article{Feng_Giglio_Xiu_2020,
  author   = {Feng, Guanhao and Giglio, Stefano and Xiu, Dacheng},
  title    = {Taming the Factor Zoo: A Test of New Factors},
  journal  = {The Journal of Finance},
  year     = {2020},
  issue    = {forthcoming},
  doi      = {10.1111/jofi.12883},
  eprint   = {https://onlinelibrary.wiley.com/doi/pdf/10.1111/jofi.12883},
  url      = {https://onlinelibrary.wiley.com/doi/abs/10.1111/jofi.12883},
}

@ARTICLE{Liu_2016,
title={Portfolio Diversification and International Corporate Bonds},
volume={51},
DOI={10.1017/S002210901600034X},
number={3},
journal={Journal of Financial and Quantitative Analysis},
publisher={Cambridge University Press},
author={Liu, Edith X.},
year={2016},
pages={959--983}
}

@Article{Kroencke_Schindler_Schrimpf_2013,
  author   = {Kroencke, Tim A. and Schindler, Felix and Schrimpf, Andreas},
  title    = {{International Diversification Benefits with Foreign Exchange Investment Styles}},
  journal  = {Review of Finance},
  year     = {2013},
  volume   = {18},
  number   = {5},
  pages    = {1847--1883},
  month    = {10},
  issn     = {1572-3097},
  doi      = {10.1093/rof/rft047},
  eprint   = {https://academic.oup.com/rof/article-pdf/18/5/1847/26316801/rft047.pdf},
  url      = {https://doi.org/10.1093/rof/rft047},
}

@ARTICLE{Barroso_SantaClara_2015,
title={Beyond the Carry Trade: Optimal Currency Portfolios},
volume={50},
DOI={10.1017/S0022109015000460},
number={5},
journal={Journal of Financial and Quantitative Analysis},
publisher={Cambridge University Press},
author={Barroso, Pedro and Santa-Clara, Pedro},
year={2015},
pages={1037--1056}
}

@ARTICLE{Ackermann_Pohl_Schmedders_2017,
author = {Ackermann, Fabian and Pohl, Walt and Schmedders, Karl},
title = {Optimal and Naive Diversification in Currency Markets},
journal = {Management Science},
volume = {63},
number = {10},
pages = {3347--3360},
year = {2017},
doi = {10.1287/mnsc.2016.2497},
URL = {https://doi.org/10.1287/mnsc.2016.2497},
eprint = {https://doi.org/10.1287/mnsc.2016.2497}
}

@ARTICLE{AddaeDapaah_Loh_2005,
author = {Addae-Dapaah, Kwame and Loh, Hwee},
title = {Exchange Rate Volatility and International Real Estate Diversification: A Comparison of Emerging and Developed Economies},
journal = {Journal of Real Estate Portfolio Management},
volume = {11},
number = {3},
pages = {225--240},
year = {2005},
doi = {10.5555/repm.11.3.l8g742127472h533},
URL = {https://aresjournals.org/doi/abs/10.5555/repm.11.3.l8g742127472h533},
eprint = {https://aresjournals.org/doi/pdf/10.5555/repm.11.3.l8g742127472h533}
}

@ARTICLE{Benjamin_Sirmans_Zietz_2001,
author = {Benjamin, John and Sirmans, Stacy and Zietz, Emily},
title = {Returns and Risk on Real Estate and Other Investments: More Evidence},
journal = {Journal of Real Estate Portfolio Management},
volume = {7},
number = {3},
pages = {183--214},
year = {2001},
doi = {10.5555/repm.7.3.b27x8337007445g5},
URL = {https://www.aresjournals.org/doi/abs/10.5555/repm.7.3.b27x8337007445g5},
eprint = {https://www.aresjournals.org/doi/pdf/10.5555/repm.7.3.b27x8337007445g5}
}

@ARTICLE{Gompers_et_al_2010,
title = "Performance persistence in entrepreneurship",
journal = "Journal of Financial Economics",
volume = "96",
number = "1",
pages = "18--32",
year = "2010",
issn = "0304-405X",
doi = "https://doi.org/10.1016/j.jfineco.2009.11.001",
url = "http://www.sciencedirect.com/science/article/pii/S0304405X09002311",
author = "Paul Gompers and Anna Kovner and Josh Lerner and David Scharfstein"
}

@ARTICLE{Hoang_et_al_2015,
title = "Is gold good for portfolio diversification? A stochastic dominance analysis of the Paris stock exchange",
journal = "International Review of Financial Analysis",
volume = "42",
pages = "98--108",
year = "2015",
issn = "1057-5219",
doi = "https://doi.org/10.1016/j.irfa.2014.11.020",
url = "http://www.sciencedirect.com/science/article/pii/S1057521915000113",
author = "Thi-Hong-Van Hoang and Hooi Hooi Lean and Wing-Keung Wong"
}

@ARTICLE{Belousova_Dorfleitner_2012,
          volume = {36},
	  issue  = {9},
          author = {Julia Belousova and Gregor Dorfleitner},
           title = {On the diversification benefits of commodities from the perspective of euro investors},
       publisher = {Elsevier},
            year = {2012},
           pages = {2455--2472},
         journal = {Journal of Banking and Finance},
             url = {https://epub.uni-regensburg.de/24267/}
}

@ARTICLE{Fogarty_2010,
title={Wine Investment and Portfolio Diversification Gains},
volume={5},
DOI={10.1017/S1931436100001401},
ngumber={1},
journal={Journal of Wine Economics},
publisher={Cambridge University Press},
author={Fogarty, James J.},
year={2010},
pages={119--131}
}

@ARTICLE{Chu_2014,
author = {Chu, Patrick},
year = {2014},
month = {02},
pages = {123--139},
title = {Study on the Diversification Ability of Fine Wine Investment},
volume = {23},
journal = {The Journal of Investing},
doi = {10.3905/joi.2014.23.1.123}
}

@Article{Momtaz_2019,
  author    = {Paul P. Momtaz},
  title     = {The Pricing and Performance of Cryptocurrency},
  journal   = {The European Journal of Finance},
  year      = {2019},
  issue     = {forthcoming},
  doi       = {10.1080/1351847X.2019.1647259},
  eprint    = {https://doi.org/10.1080/1351847X.2019.1647259},
  publisher = {Routledge},
  url       = {https://doi.org/10.1080/1351847X.2019.1647259},
}

@ARTICLE{Adhami_Guidici_Martinazzi_2018,
title = "Why do businesses go crypto? An empirical analysis of initial coin offerings",
journal = "Journal of Economics and Business",
volume = "100",
pages = "64--75",
year = "2018",
note = "FinTech -- Impact on Consumers, Banking and Regulatory Policy",
issn = "0148-6195",
doi = "https://doi.org/10.1016/j.jeconbus.2018.04.001",
url = "http://www.sciencedirect.com/science/article/pii/S0148619517302308",
author = "Saman Adhami and Giancarlo Giudici and Stefano Martinazzi"
}

@InCollection{Vayanos_et_al_2013,
  author    = {Dimitri Vayanos and Jiang Wang},
  title     = {Market Liquidity---Theory and Empirical Evidence},
  booktitle = {Handbook of the Economics of Finance},
  publisher = {Elsevier},
  year      = {2013},
  editor    = {George M.\ Constantinides and Milton Harris and Rene M.\ Stulz},
  volume    = {2},
  chapter   = {19},
  pages     = {1289--1361},
  doi       = {https://doi.org/10.1016/B978-0-44-459406-8.00019-6},
  issn      = {1574-0102},
  url       = {http://www.sciencedirect.com/science/article/pii/B9780444594068000196},
}

@ARTICLE{Scharnowski_2020,
title = "Understanding Bitcoin liquidity",
journal = "Finance Research Letters",
pages = "101477",
year = "2020",
issn = "1544-6123",
doi = "https://doi.org/10.1016/j.frl.2020.101477",
url = "http://www.sciencedirect.com/science/article/pii/S1544612319311286",
author = "Stefan Scharnowski"
}

@ARTICLE{Brauneis_et_al_2020,
title = "What Drives the Liquidity of Cryptocurrencies? A Long-Term Analysis",
journal = "Finance Research Letters",
pages = "101537",
year = "2020",
issn = "1544-6123",
doi = "https://doi.org/10.1016/j.frl.2020.101537",
url = "http://www.sciencedirect.com/science/article/pii/S154461231931400X",
author = "Alexander Brauneis and Roland Mestel and Erik Theissen"
}

@ARTICLE{Mehra_Prescott_1985,
title = "The equity premium: {A} puzzle",
journal = "Journal of Monetary Economics",
volume = "15",
number = "2",
pages = "145--161",
year = "1985",
issn = "0304-3932",
doi = "https://doi.org/10.1016/0304-3932(85)90061-3",
url = "http://www.sciencedirect.com/science/article/pii/0304393285900613",
author = "Rajnish Mehra and Edward C.\ Prescott"
}

@ARTICLE{Mei_Moses_2002-AER,
Author = {Mei, Jianping and Moses, Michael},
Title = {Art as an Investment and the Underperformance of Masterpieces      },
Journal = {American Economic Review},
Volume = {92},
Number = {5},
Year = {2002},
Month = {12},
Pages = {1656--1668},
DOI = {10.1257/000282802762024719},
URL = {https://www.aeaweb.org/articles?id=10.1257/000282802762024719}}

@ARTICLE{Campbell_2008,
	author = {Campbell, Rachel.},
	title = {Art as a Financial Investment},
	volume = {10},
	number = {4},
	pages = {64--81},
	year = {2008},
	doi = {10.3905/jai.2008.705533},
	publisher = {Institutional Investor Journals Umbrella},
	issn = {1520-3255},
	URL = {https://jai.pm-research.com/content/10/4/64},
	eprint = {https://jai.pm-research.com/content/10/4/64.full.pdf},
	journal = {The Journal of Alternative Investments}
}

@Article{Amihud2002,
  author    = {Amihud, Yakov},
  title     = {Illiquidity and stock returns: cross-section and time-series effects},
  journal   = {Journal of financial markets},
  year      = {2002},
  volume    = {5},
  number    = {1},
  pages     = {31--56},
  publisher = {Elsevier},
}

@UNPUBLISHED{Pele_et_al_2020,
  author = {Pele, Daniel Traian and Wesselh\"offt, Niels and H\"ardle, Wolfgang K. and Kolossiatis, Michalis and Yatracos, Yannis G.},
  title  = {A Statistical Classification of Cryptocurrencies},
  year   = {2020},
  type   = {{SSRN} {Scholarly} {Paper}},
  number = {ID 3548462},
  address= {Rochester, NY},
  note   = {Available at SSRN 3548462},
}
